# Long-term Orbital Period Variation of Hot Jupiters from Transiting Time Analysis using TESS Survey Data

Wenqin Wang,[1,2] Zixin Zhang,[1,2] Zhangliang Chen,[1,2] Yonghao Wang,[3] Cong Yu,[1,2] and Bo Ma[1,2]

[1]*School of Physics and Astronomy, Sun Yat-sen University, Zhuhai 519082, China; mabo8@mail.sysu.edu.cn*
[2]*CSST Science Center for the Guangdong-Hong Kong-Macau Great Bay Area, Sun Yat-sen University, Zhuhai 519082, China*
[3]*School of Science, Hainan University, Haikou 570228, China*

## ABSTRACT

Many hot Jupiters may experience orbital decays, which are manifested as long-term transit timing variations. We have analyzed 7068 transits from the Transiting Exoplanet Survey Satellite (TESS) for a sample of 326 hot Jupiters. These new mid-transit time data allow us to update ephemerides for these systems. By combining the new TESS transit timing data with archival data, we search for possible long-term orbital period variations in these hot Jupiters using a linear and a quadratic ephemeris model. We identified 26 candidates that exhibit possible long-term orbital period variations, including 18 candidates with decreasing orbital periods and 8 candidates with increasing orbital periods. Among them, 12 candidates have failed in our leave-one-out cross-validation (LOOCV) test and thus should be considered as marginal candidates. In addition to tidal interaction, alternative mechanisms such as apsidal precession, Rømer effect, and Applegate effect could also contribute to the observed period variations. The ephemerides derived in this work are useful for scheduling follow-up observations for these hot Jupiters in the future. The Python code (PdotQuest) used to generate the ephemerides is made available online.

## 1. INTRODUCTION

Hot Jupiters (HJs) are a type of exoplanet with masses comparable to Jupiter and orbital periods shorter than ten days. As a result, they can be readily detected through ground-based transit surveys and have been the subject of extensive long-term investigations. Due to their close proximity to the host stars, they are expected to undergo intense tidal interactions with the stars (Ogilvie 2014) . The tidal bulges induced by hot Jupiters on their host stars generate torques that can transfer angular momentum from the planets to the stars (Rasio et al. 1996; Levrard et al. 2009), known as equilibrium tides (Counselman 1973; Rasio et al. 1996). Particularly, when the orbital period of a hot

Corresponding author: Bo Ma
mabo8@mail.sysu.edu.cn

Jupiter is shorter than the rotation period of its host star (Penev et al. 2018) , the star may experience spin-up, and the hot Jupiter may spiral inward over time, potentially leading to tidal disruption or destruction by Roche lobe overflow or atmospheric evaporation (Levrard et al. 2009; Jackson et al. 2009; Matsumura et al. 2010). The decay rates can vary depending on the magnitude of the stellar tidal dissipation for a given configuration (Levrard et al. 2009; Matsumura et al. 2010). Population studies offer additional evidences for orbital decay, such as the scarcity of gaseous giants with periods less than one day (Jackson et al. 2008; Hansen 2010; Penev et al. 2012; Ogilvie 2014), the unusually rapid rotation of certain hot Jupiters' host stars (Penev et al. 2018), and the rarity of hot Jupiters around subgiants (Hansen 2010; Schlaufman & Winn 2013).

The orbital decay of hot Jupiters can be revealed through monitoring of their transit timings over decades, known as transit timing variations (TTVs). Currently, the detection of orbital period decay has been confirmed in the WASP-12 system through various TTV studies (Maciejewski et al. 2016a; Patra et al. 2017; Yee et al. 2020). Another candidate is WASP-4 b (Bouma et al. 2019), although the evidence is less compelling. Long-term TTVs of hot Jupiters can also be induced by other physical mechanisms, such as planetary mass loss (Valsecchi et al. 2015; Jackson et al. 2016), apsidal precession (Miralda-Escudé 2002; Ragozzine & Wolf 2009), the Rømer effect, i.e. the line-of-sight acceleration due to wide stellar companions (Bouma et al. 2019), and the Applegate mechanism (Applegate 1992). Short-term TTVs can be used to detect extra planets in the same planetary system (Holman & Murray 2005; Agol et al. 2005). Additionally, overestimation of the measurement precision of transit time data could also introduce false TTV signal (Mallonn et al. 2019).

To test the various scenarios of the long-term orbital period variations, a large and precise follow-up transit observation dataset is required. The Transiting Exoplanet Survey Satellite (TESS; Ricker et al. 2014), launched in 2018, has provided such a great opportunity. It offers the latest and precise transit timing measurements for a lot of known hot Jupiters, which are suitable for long-term TTV studies. By combining the high-precision 2-minute cadence transit data provided by TESS with archival data from previous work, new constraints can be placed on the period change rate of these hot Jupiters and the tidal dissipation factor of their host stars. Furthermore, transit ephemeris updates provided by TESS can significantly enhance our capability to predict the future transit times of hot Jupiters, which is crucial for future space telescopes targeting these hot Jupiters.

The paper is organized as follows. Section 2 describes our sample selection and timing analysis. Section 3 introduces the transit timing models and the model fitting processes. Section 4 presents our results, including the candidate systems exhibiting period decay, period increase, and constant period. In Section 5 and 6, we summarize our main findings, compare our study with previous studies, and discuss the possible physical origin of the observed long-term timing variations.

## 2. SAMPLE SELECTION AND TESS DATA ANALYSIS

In this study, we select transiting hot Jupiters (with an orbital period less than 10 days and the mass larger than 0.3 $M_J$) observed by TESS. We also require the planet has previous transiting data before the TESS mission, thus providing a longer time baseline, which is crucial for detecting long term orbital period variations. After a rigorous searching effort, we find a total of 326 hot Jupiters in our whole sample.

We then download the 2-minute cadence Presearch Data Conditioning-Simple Aperture Photometry (PDC-SAP) TESS light curves(Team 2021) from the MAST website. Light curves showing excessive noises are excluded from the following analysis. We fit a Mandel & Agol (2002) model to the phase-folded light curve of each TESS sector. The parameters used for the transit model included zero epoch $T_0$, period $P$, the impact parameter $b$, stellar density in $g/cm^3$, and planet-to-star radius ratio $R_p/R_*$. The eccentricity and argument of periastron were fitted using the parameterization of $\sqrt{e}cos\omega$ and $\sqrt{e}sin\omega$. In addition, two quadratic limb darkening parameters were fit for each band. For each transiting hot Jupiter, we fit first the combined light curve before fit for individual transits, with an example of WASP-12 b shown in Figure 1 and 3. In Figure 1 we show the phase folded light curve of WASP-12 b and the best-fit transit light curve model, while in Figure 2 we show all the individual transit light curves of the first sector of TESS data for WASP-12 b. The derived all TESS transit times for WASP-12 b are displayed in Figure 3. The epoch number is calculated relative to a reference zero epoch from literature. For our sample of 326 hot Jupiters, we obtain a total of 7,068 TESS light curves and middle transit times (Table 1).

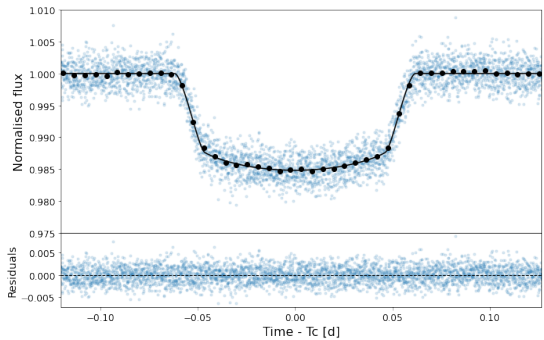

**Figure 1.** The phase folded light curve of WASP-12 b from TESS. The best-fit transit light curve model from Mandel & Agol (2002) is also shown in black solid line.

## 3. MODELING TRANSIT TIMES DATA

In this section, we use literature transit times, which are mainly from the database of Ivshina & Winn (2022), and TESS transit times from this work to search for possible long term orbital decay for our sample of 326 hot Jupiters. We also include some high quality light curves

**Table 1.** Transit times.

| System | Epoch | $T_c(\mathrm{BJD_{TDB}})$ | Uncertainty (days) |
|---|---|---|---|
| CoRoT-1 b | 1458 | 2458469.06773 | 0.00085 |

NOTE—Only a portion of this table is displayed here to illustrate its format and content. The complete, machine-readable version of the table can be accessed in the online version of this work.

from the Exoplanet Transit Database (ETD; Poddaný et al. 2010) in our analysis when necessary.

### 3.1. *Two Transit-timing Models*

We fit two models to the transit timing data. The first model assumes a constant orbital period and hence a linear ephemeris:

$$T_N = T_0 + NP, \qquad (1)$$

where $T_N$, $T_0$, N and $P$ are the calculated mid-transit times, the reference mid-transit time, the number of orbits counted from the designated reference transit, and the orbital period of the planet respectively. The free parameters $T_0$ and $P$ are to be fitted in this model, with initial guess values taken from discovery paper.

In the second model, we assume the planet has a constant rate of change of orbital angular momentum, which will cause a constant period change rate $a = \frac{dP}{dt} = \dot{P}$, usually in the unit of ms/yr. The corresponding ephemeris model can be calculated using the following two equations:

$$T_N = T_{N-1} + P_{N-1} \qquad (2)$$
$$P_N = P_{N-1} + a(P_{N-1} + P_N)/2, \qquad (3)$$

where $T_N$, $T_0$, $N$, $P_0$, and $a$ are the calculated mid-transit times, the reference mid-transit time, the number of orbits counted from the designated reference transit, the initial orbital period of the planet, and the constant period change rate, respectively. Eqn.(3) means each succesive orbital period is slightly different from its previous one. Notice that the orbital period change rate $a$ can also be expressed as $a = \frac{dP}{dt} = \frac{1}{P}\frac{dP}{dN}$, which can be used to convert our equations to the quadratic equation form used in Maciejewski et al. (2021). The three free parameters to be fitted are the reference epoch $T_0$, the initial period $P_0$ at the reference epoch, and the constant period change rate $a$, with initial guesses for the first two taken from literature values.

### 3.2. *A New Tool for Long-term Transit Timing Data Analysis*

We have developed a new software tool, PdotQuest, to analyze the long-term transit timing variations using both of a linear ephemeris model and a quadratic ephemeris model. This software allows for the efficient fitting of data by simply inputting a column of middle transit times, a column of transit time measurement uncertainties, as well as an initial orbital period corresponding to the earliest transit time. The best-fit results for the two ephemeris models can then be displayed and compared.

In the software, we employ the emcee package (Foreman-Mackey et al. 2013) to determine the optimal model parameters by minimizing the $\chi^2$ statistic and calculate uncertainties of all fitting parameters. We run the MCMC sampling with 100 walkers and burn the initial 20% of the 500 steps used for each walker to ensure convergence. Broad uniform priors are applied to all fitting parameters.

A "n-sigma rejection" iterative fitting scheme has been employed for data clipping to remove outliers during the fitting process. The standard deviation of the fitting residuals is calculated, and any data point with residual value outside of a n-$\sigma$ range from the residual mean is eliminated. After a few trial, we find the "3-sigma rejection" scheme sometimes removes useful data points from the fitting and the "5-sigma rejection" scheme sometimes keeps too many outliers. Thus, we decide to utilize both of the "3-sigma rejection" and the "5-sigma rejection" clipping strategy during our fitting, and reach a conclusion based on the two fitting results.

The Bayesian Information Criterion (BIC) is utilized to compare the relative quality of each model in describing the transit timing data, given the different number of free parameters among the models. The BIC is defined as:

$$BIC = \chi^2 + k \log n, \qquad (4)$$

where n is the sample size, and k is the number of free parameters in the model. A lower BIC score usually signals a better model, with $\Delta BIC > 10$ corresponding to 'very strong' evidence in favor of the model with smaller BIC (Kass & Raftery 1995). In our analysis, we select hot Jupiter candidates showing signs of long-term period change using the following two criteria: (a) the constant period change rate $a$ is at least 3-$\sigma$ away from zero value, and (b) $\Delta BIC > 10$ where $\Delta BIC$ is the BIC difference between the linear and quadratic model fitting.

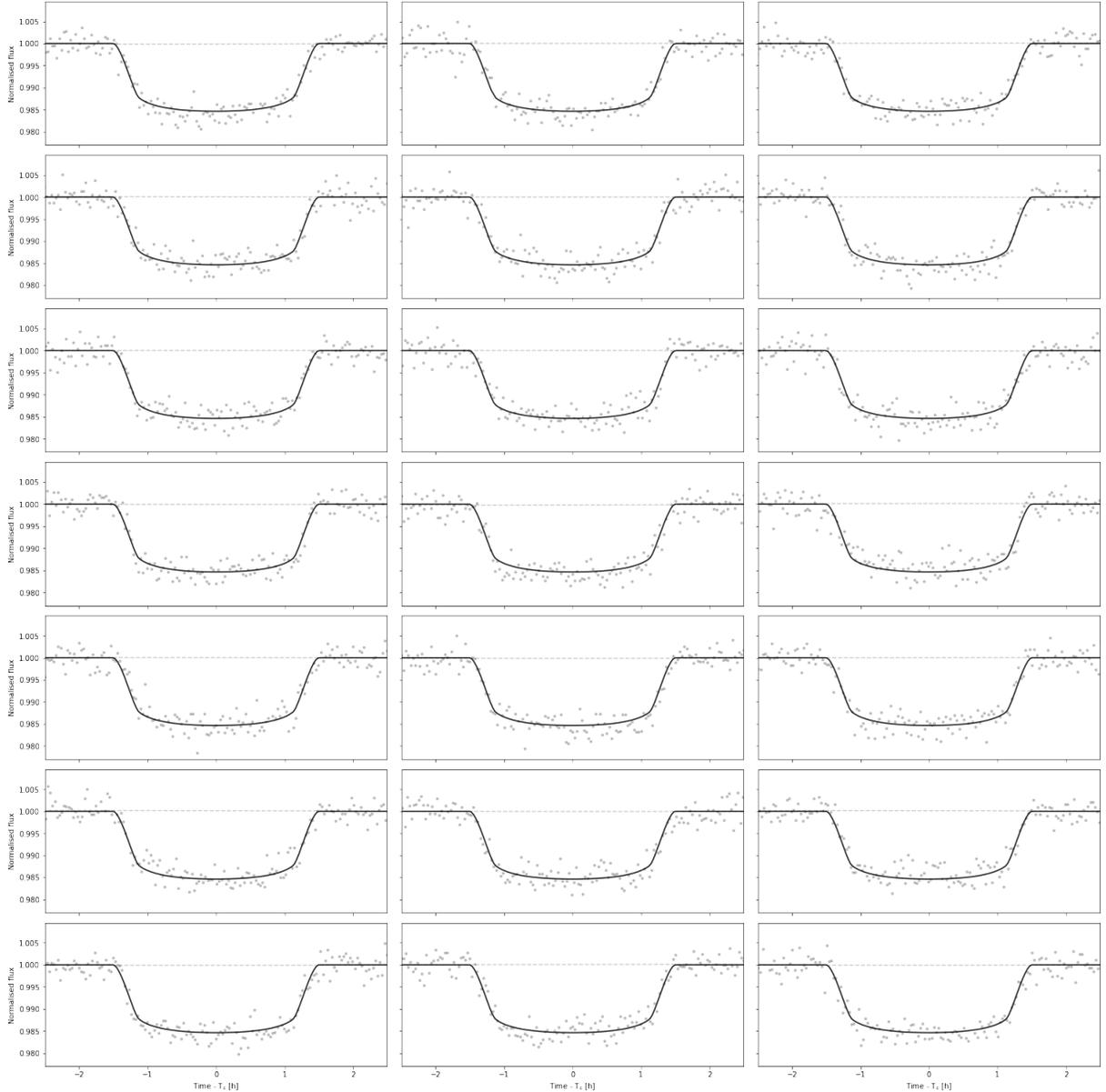

**Figure 2.** Individual transit light curve of WASP-12b from sector 1 observations of TESS.

For all of our long-term period variation candidates, we also introduce a new test called "leave-one-out cross-validation" (LOOCV), in which we remove one transit data point at a time and re-fit the remaining data points. In this way, we are able to identify the most influential data points and assess the robustness of the quadratic model fitting against the removal of any individual data point. This new test is motivated by the fact that some transit time data from literature have unrealistic small error bars and can yield significantly biased fitting results, with HD 189733 b as an example shown in Sec 4.

## 4. RESULTS

In this section, we present the fitting results for each hot Jupiter analyzed from our sample. We divide the whole hot Jupiter sample into three different categories: 18 hot Jupiters showing signs of orbital period decay, 8 hot Jupiters showing signs of orbital period increase, and 300 hot Jupiters showing no signs of period change. All the ephemerides derived in this work are made available online (Table 2).

**Table 2.** Ephemerides.

| System | $T_0$(BJD$_{\mathrm{TDB}}$) | Uncertainty (days) | $P_0$(days) | Uncertainty (days) | $\dot{P}$ (ms/yr) | $\sigma$ rejction |
|---|---|---|---|---|---|---|
| CoRoT-1 b | 2454138.32729 | 0.00005 | 1.50896863 | 0.00000005 | / | / |
| CoRoT-2 b | 2453566.47840 | 0.00028 | 1.74300015 | 0.00000035 | -21.65 ± 2.96 | 5 |
| CoRoT-2 b | 2453566.47894 | 0.00028 | 1.74299976 | 0.00000034 | -19.12 ± 2.90 | 3 |

NOTE—The table contains ephemerides for both samples fitted by a linear model and TTV candidates fitted by a quadratic model. The ephemerides of TTV candidates are presented on two separate lines as they were fitted using different sigma-rejection schemes. Only a portion of this table is displayed here to illustrate its format and content. The complete, machine-readable version of the table can be accessed in the online version of this work.

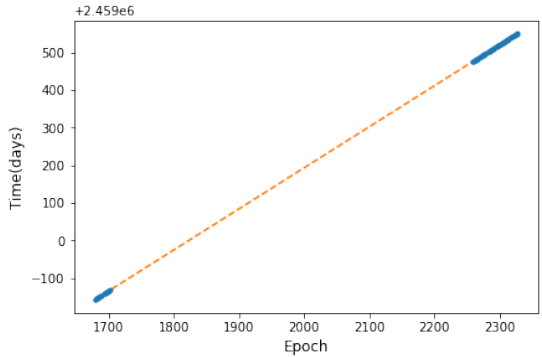

**Figure 3.** TESS transit times for WASP-12 b. The epoch number is calculated relative to an reference zero epoch from literature.

## 4.1. Candidates with Decreasing Orbital Periods

We have 18 hot Jupiters showing signs of orbital period decay. Among them, the last 7 (from 4.1.12 to 4.1.18) have failed to pass the LOOCV test, i.e. upon the removal of one single data point, they no longer meet the two criteria we set above for identifying long-term period variation candidates. .

### 4.1.1. WASP-12 b

WASP-12 b is an ultra-hot Jupiter first reported by Hebb et al. (2009), with a mass of 1.47 $M_J$ and radius of 1.9 $R_J$. It orbits a 6150 K late F-type star with a period of 1.091 days. The host star properties are consistent with a 1.3$M_\odot$ main sequence star or a 1.2$M_\odot$ subgiant without a convective core (Weinberg et al. 2017). Its decreasing orbital period was first detected by Maciejewski et al. (2016a), and subsequent studies have confirmed the period change (Patra et al. 2017; Baluev et al. 2019) and established orbital decay as its cause using transit and occultation observations (Yee et al. 2020; Turner et al. 2021). The most recent research conducted by Wong et al. (2022) revealed a decay rate of $-29.81 \pm 0.94$ ms/yr, while Hagey et al. (2022) obtained a result of $-29.1 \pm 1.0$ ms/yr based on the ETD transit times. We have analyzed 189 mid-transit time data for this system (107 from literature, 82 from TESS). Our best-fitting result is $\dot{P} = -30.19 \pm 0.92$ ms/yr, with the timing residual O-C diagram shown in Figure 4. The downward parabola in the plot clearly reveals the secular period decrease trend. WASP-12 b still remains as the best orbital period decay candidates.

### 4.1.2. WASP-4 b

WASP-4 b is a hot Jupiter first identified by Wilson et al. (2008), with a mass of 1.2 $M_J$ and radius of 1.4 $R_J$. It orbits a G7V star on a 1.338-day orbit. Like WASP-12 b, WASP-4 b is also a well-studied planet showing signs of orbital decay. Bouma et al. (2019) first reported the orbital variation of WASP-4 b to be $-12.6 \pm 1.2$ ms/yr using TESS and ground-based observations, and Southworth et al. (2019) found a period decay rate of $-9.2 \pm 1.1$ ms/yr using only ground-based observations. Baluev et al. (2020) have re-analyzed 124 transit light curves of WASP-4 b and obtained a period derivative of $-5.94 \pm 0.39$ ms/yr. They have also analyzed radial velocity (RV) data and attributed the orbital period variation to the line-of-sight acceleration caused by a wide-orbit companion. Recently, a comprehensive analysis conducted by Turner et al. (2022) does not find acceleration in the RV data, and suggests the possible existence of another wide-orbit high-mass planet WASP-4 c in this system. They tend to use the orbital decay scenario to explain the TTV seen in WASP-4 b, with a rate of $-7.33 \pm 0.71$ ms/yr. We have analyzed 122 mid-transit time data (75 from literature, 47 from TESS). Our analysis yields a period change rate of $\dot{P} = -6.43 \pm 0.55$ ms/yr (Figure A1) , which is consistent with the $-5.81 \pm 1.58$ ms/yr value reported by Ivshina & Winn (2022) and the $-4.8 \pm 1.4$ ms/yr value reported by Maciejewski et al. (2022) but with a smaller error bar.

**Table 3.** Model fitting and statistical results for 26 hot Jupiters

| System | $\dot{P}$ (ms/yr) | $\text{BIC}_{linear}$ | $\text{BIC}_{quad}$ | $\dot{P}$ (ms/yr) | $\text{BIC}_{linear}$ | $\text{BIC}_{quad}$ | LOOCV |
|--------|-------------------|----------------------|---------------------|-------------------|----------------------|---------------------|-------|
| | (5-$\sigma$ rejction) | | | (3-$\sigma$ rejction) | | | |
| CoRoT-2 b | -21.65 ± 2.96 | 634.56 | 579.03 | -19.12 ± 2.90 | 332.00 | 274.41 | pass |
| HD189733 b | -8.44 ± 2.44 | 58.98 | 47.33 | -8.69 ± 2.42 | 41.95 | 25.08 | fail |
| HAT-P-37 b | -34.29 ± 6.67 | 221.71 | 195.71 | -34.03 ± 6.74 | 221.71 | 195.71 | pass |
| KELT-16 b | -27.82 ± 4.76 | 103.63 | 70.77 | -26.64 ± 4.80 | 86.95 | 56.42 | pass |
| TrES-1 b | -14.82 ± 2.37 | 355.98 | 321.61 | -14.89 ± 2.35 | 249.19 | 203.68 | pass |
| TrES-3 b | -8.77 ± 0.47 | 1834.16 | 1510.6 | -8.51 ± 0.48 | 1695.13 | 1377.28 | fail |
| TrES-5 b | -9.71 ± 2.11 | 216.66 | 194.11 | -8.97 ± 2.06 | 169.19 | 153.79 | pass |
| WASP-4 b | -6.43 ± 0.56 | 342.39 | 213.62 | -6.47 ± 0.58 | 276.84 | 160.39 | pass |
| WASP-10 b | -27.74 ± 3.49 | 186.51 | 123.57 | -28.03 ± 3.65 | 180.41 | 116.80 | pass |
| WASP-12 b | -30.19 ± 0.92 | 1256.13 | 222.20 | -30.37 ± 0.95 | 1101.08 | 216.94 | pass |
| WASP-16 b | -51.98 ± 12.87 | 99.24 | 84.17 | -53.33 ± 13.99 | 94.21 | 78.96 | fail |
| WASP-19 b | -1.64 ± 0.29 | 693.77 | 661.02 | -1.59 ± 0.29 | 618.58 | 604.08 | fail |
| WASP-22 b | -71.76 ± 17.12 | 38.79 | 21.64 | -69.15 ± 16.86 | 35.91 | 19.44 | pass |
| WASP-45 b | -169.21 ± 21.09 | 350.59 | 289.88 | -166.11 ± 21.67 | 344.01 | 285.29 | pass |
| WASP-47 b | -48.45 ± 14.14 | 77.18 | 65.36 | -49.98 ± 14.15 | 72.24 | 59.76 | fail |
| WASP-80 b | -23.04 ± 5.15 | 4017.15 | 3996.25 | -31.99 ± 4.91 | 343.96 | 301.65 | fail |
| XO-3 b | -31.59 ± 5.24 | 350.35 | 313.85 | -27.69 ± 5.29 | 310.94 | 276.60 | fail |
| XO-4 b | -62.57 ± 17.48 | 90.84 | 77.61 | -61.16 ± 17.38 | 90.84 | 77.62 | pass |
| HAT-P-7 b | 18.28 ± 4.05 | 112.96 | 92.35 | 20.79 ± 4.31 | 52.18 | 32.16 | pass |
| HAT-P-43 b | 94.40 ± 22.55 | 70.24 | 53.24 | 108.34 ± 23.91 | 63.60 | 33.84 | fail |
| HAT-P-44 b | 119.67 ± 25.89 | 63.69 | 42.49 | 118.92 ± 25.89 | 63.69 | 42.49 | fail |
| WASP-1 b | 22.50 ± 5.23 | 88.21 | 69.68 | 24.33 ± 5.19 | 60.07 | 38.75 | pass |
| WASP-6 b | 20.66 ± 5.19 | 32.84 | 17.30 | 20.80 ± 5.15 | 32.84 | 17.30 | fail |
| WASP-11 b | 23.32 ± 6.69 | 102.65 | 89.70 | 23.44 ± 6.59 | 102.65 | 89.70 | fail |
| WASP-17 b | 77.64 ± 8.19 | 550.63 | 457.85 | 77.47 ± 8.13 | 550.63 | 457.85 | fail |
| WASP-46 b | 51.68 ± 2.86 | 2852.85 | 2512.55 | 63.22 ± 2.78 | 2092.71 | 1305.92 | pass |

### 4.1.3. CoRoT-2 b

CoRoT-2 b is a hot Jupiter discovered by Alonso et al. (2008), with a mass of 3.3 $M_J$ and radius of 1.5 $R_J$. It orbits an active G7V star on a 1.743-day orbit. Since its discovery, CoRoT-2 b has been the subject of numerous studies, including observations of its atmosphere and studies of its orbital dynamics. The study of Ivshina & Winn (2022) gave a period change rate of $\dot{P} = -103.76 \pm 6.33$ ms/yr, where they have used data from Öztürk & Erdem (2019). We notice there exhibits a significant disparity between the error bars of transit times ranging from 2454237.53562 to 2454378.7143 JD (Öztürk & Erdem 2019) and the fitting residuals, which is shown in the top panel of Figure 5. For the purpose of assigning realistic uncertainties to the data, we have employed a formula $e_{new} = \sqrt{(\sigma_{resi}^2 + \Sigma e^2)}$ to manually inflate the error bars of the data points shown in Figure 5, where $\sigma_{resi}$ is the standard deviation and $e$ is the original measurement uncertainty. Comparison of the transit timing data before and after the uncertainty inflation is shown on Figure 5. We have analyzed 123 mid-transit time data (118 from literature, 5 from TESS). With the inclusion of TESS data, our best-fitting result is now $\dot{P} = -21.65 \pm 2.96$ ms/yr using the 5-$\sigma$ rejection scheme and $-19.12 \pm 2.90$ ms/yr using the 3-$\sigma$ rejection scheme (Figure A2) , much smaller than the result from Ivshina & Winn (2022).

### 4.1.4. HAT-P-37 b

HAT-P-37 b is a 1.2 $M_J$, 1.2 $R_J$ hot Jupiter discovered by Bakos et al. (2012). It orbits a G-type star with a period of 2.797 days. Baluev et al. (2019) homogenously analyzed the light curves of HAT-P-37 b and found a period change rate of $-27.2 \pm 8.8$ ms/yr, which showed a slight preference for the orbital decay model over a constant period model. A-thano et al. (2022) has reported a period change rate of $-80 \pm 30$ ms/yr by using ground-based observation. However, the corresponding $Q_*'$ is too small and inconsistent with theoretical estimation. As a result, they favor the apsidal precession model and suggest the TTV can be explained by light-time effect (LiTE). TESS has observed HAT-P-37 b in seven sectors, but the first six cannot be used due to significant background contamination. We have analyzed 35 mid-transit time data (27 from literature, 8 from TESS). Upon inclusion of the only usable sector of TESS data, our best-fitting orbital decay rate is $\dot{P} = -34.29 \pm 6.67$ ms/yr (Figure A3).

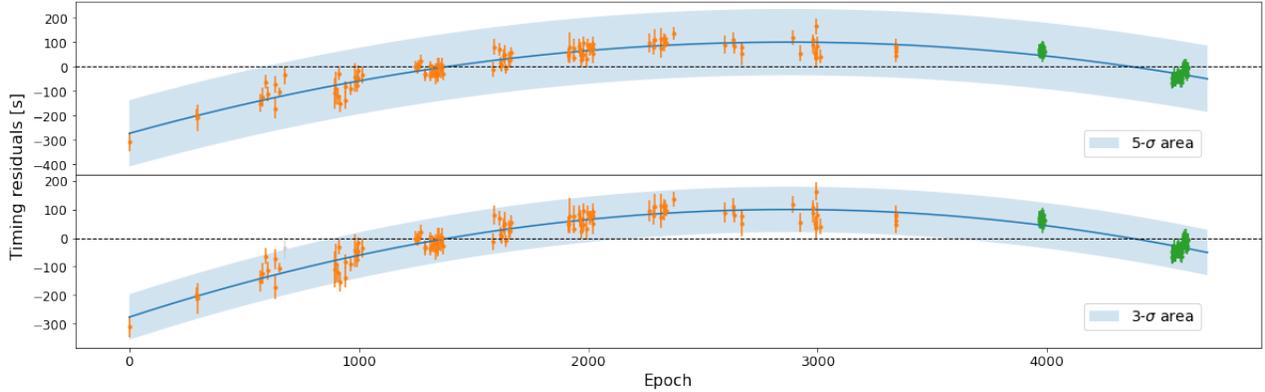

**Figure 4.** Timing residuals of WASP-12 b. The top panel displays the fitting results using a 5-σ rejection scheme, while the bottom panel shows the fitting results using a 3-σ rejection scheme. The blue curves and shaded areas indicate the best-fit quadratic model and corresponding 5 − σ (top) and 3 − σ (bottom) confidence regions. The orange points are based on literature data. The green points are based on TESS data. The gray points are clipped data.

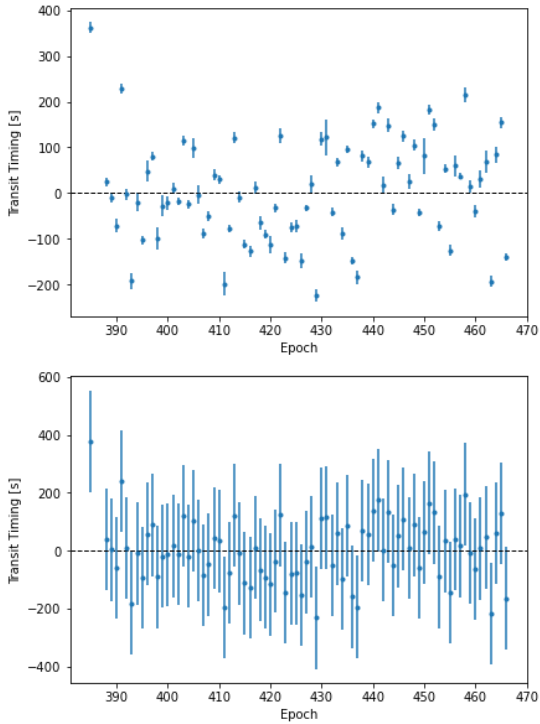

**Figure 5.** Part of the transit timing residuals of CoRoT-2 b. The top panel presents the raw data, which are taken from Öztürk & Erdem (2019), while the bottom panel displays the same data with inflated error bars. In both panels, we show the residuals of the linear model fit to provide a clearer visual comparison, where it is clearly seen that the original measurement uncertainties from Öztürk & Erdem (2019) are too small.

### 4.1.5. *KELT-16 b*

KELT-16 b is a highly irradiated, ultra-short period hot Jupiter discovered by Oberst et al. (2017), with a mass of 2.75 $M_J$ and a radius of 1.4 $R_J$. It orbits an F7V star with a period of 0.97 days. We have analyzed 111 mid-transit time data (46 from literature, 65 from TESS). Our best-fitting value of $\dot{P} = -27.82 \pm 4.76$ ms/yr (Figure A4) is different from previous studies, which all agree on a constant orbital period model (Maciejewski et al. 2018; Patra et al. 2020; Mancini et al. 2022). Given the relatively short time span since the planet's discovery, we anticipate that more observation data will be gathered in the future to refine our understanding of this system.

### 4.1.6. *TrES-1 b*

TrES-1 b is a 0.8 $M_J$, 1.1 $R_J$ hot Jupiter discovered by the Trans-Atlantic Exoplanet Survey (Alonso et al. 2004). It orbits a K0V star with a period of 3.03 days. We have analyzed 75 mid-transit time data (41 from literature, 34 from TESS). Our model fitting analysis reveals a $\dot{P}$ of $-14.82 \pm 2.37$ ms/yr (Figure A5) , which is consistent with the $-18.36 \pm 3.73$ ms/yr value from Ivshina & Winn (2022) and $-10.9 \pm 2.1$ ms/yr value from Hagey et al. (2022).

### 4.1.7. *TrES-5 b*

TrES-5 b is a 1.8 $M_J$, 1.2 $R_J$ hot Jupiter discovered by Mandushev et al. (2011). It orbits a K-type star with a period of 1.482 days. Sokov et al. (2018) suggested the existence of a second planet in the system on a 1:2 resonance orbit, with a mass of 0.24 $M_J$ based from a 99-day period TTV analysis. Maciejewski et al. (2021) could not confirm the presence of an additional planet but found a long-term variation of the orbital period of TrES-5b at a rate of $-20.4 \pm 4.7$ ms/yr. They suggested this variation is caused by a line-of-sight acceleration of the system induced by a massive wide-orbit companion. Ivshina & Winn (2022) analyzed TESS data and found the period to be changing at a rate of $-17.47 \pm 3.79$ ms/yr. We have analyzed 121 mid-transit time data (55 from literature, 66 from TESS). Our analysis yields a period decay rate of $-9.71 \pm 2.11$ ms/yr based on three additional sectors of TESS data (Figure A6). TrES-5 b is an intriguing target that warrants further observations.

### 4.1.8. WASP-10 b

WASP-10 b is a 3.2 $M_J$, 1.1 $R_J$ hot Jupiter discovered by Christian et al. (2009). It orbits a K5V star with a period of 3.093 days. Previous studies have investigated the short-term TTVs of WASP-10 b, with proposed causes including starspot occultations (Barros et al. 2013) or a 0.1 $M_J$ companion with a 5.23-day period (Maciejewski et al. 2011), which remains unconfirmed. RV analysis of Knutson et al. (2014) suggested the presence of a potential massive distant companion. Using the ETD data, Hagey et al. (2022) derived a best-fitting decay rate of $-21.9 \pm 2.4$ ms/yr. We have analyzed 49 mid-transit time data (41 from literature, 8 from TESS). Our analysis finds a period change rate of $\dot{P} = -27.74 \pm 3.49$ ms/yr (Figure A7), which is consistent with the result from Hagey et al. (2022).

### 4.1.9. WASP-22 b

WASP-22 b is a 0.6 $M_J$, 1.2 $R_J$ low-density hot Jupiter discovered by Maxted et al. (2010). It orbits a G1-type star with a period of 3.533 days. Knutson et al. (2014) confirmed its RV trend and provided evidence for the presence of a third body in the system, which could be a second planet or a M-dwarf. There is no discussion of its long-term TTV trends from previous studies. We have analyzed 16 mid-transit time data (11 from literature, 5 from TESS). our best-fitting result is $\dot{P} = -71.76 \pm 17.12$ ms/yr (Figure A8). Such a large amplitude of $\dot{P}$ should be easy to verify by more observations in the future.

### 4.1.10. WASP-45 b

WASP-45 b is a 1.0 $M_J$, 1.1 $R_J$ hot Jupiter discovered by Anderson et al. (2012). It orbits a K2V star with a period of 3.126 days. Recent timing study of WASP-45 b is from Ivshina & Winn (2022), which found a period decay rate of $-262.57 \pm 28.35$ ms/yr. They labeled it as a mediocre candidate due to its early scattered data. We have analyzed 24 mid-transit time data (10 from literature, 14 from TESS). Our model fitting gives a best-fitting $\dot{P}$ of $-169.21 \pm 21.09$ ms/yr (Figure A9), which is slight smaller than the value reported by Ivshina & Winn (2022).

### 4.1.11. XO-4 b

XO-4 b is a 1.6 $M_J$, 1.3 $R_J$ hot Jupiter discovered by McCullough et al. (2008). It orbits an F5V star with a period of 4.125 days. There have been no previous reports of TTV from this planet. We have analyzed 37 mid-transit time data (21 from literature, 16 from TESS). Combining the archival data and three more sectors of TESS data, we find a best-fitting period change rate of $\dot{P} = -62.57 \pm 17.48$ ms/yr (Figure A10).

### 4.1.12. HD 189733 b

HD 189733 b was discovered by Bouchy et al. (2005), with a mass of 1.1 $M_J$ and a radius of 1.1 $R_J$. It orbits a chromospherically active K1.5 V star with a period of 2.219 days. It is one of the most extensively studied hot Jupiters due to its close proximity to the Earth. Dowling Jones et al. (2018) reported a detection of period decay with $\dot{P} = -18.8 \pm 4.3$ ms/yr. Most of their mid transit times were from ETD. We have analyzed 34 mid-transit time data (14 from literature, 20 from TESS). We find $\dot{P} = -8.44 \pm 2.44$ ms/yr, with the best-fit result shown in Figure 6. During the LOOCV analysis, we have sequentially removed each data point from the dataset and re-fitted the model accordingly. We find that for three particular data points taken from Morvan et al. (2020), the resulting fits no longer satisfy the criteria for significant orbital variation (see Figure 7 for details). The LOOCV results suggest the existence of long-term orbital decay found by Dowling Jones et al. (2018) is questionable. Further investigation with additional data is necessary to verify this conclusion.

### 4.1.13. TrES-3 b

TrES-3 b is a 1.9 $M_J$, 1.3 $R_J$ hot Jupiter discovered by O'Donovan et al. (2007). It orbits a G-type star in a 1.306-day period. Earlier studies have shown that TrES-3 b exhibits almost no orbital period change (Mannaday et al. 2020, 2022). We have analyzed 283 mid-transit time data (190 from literature, 94 from TESS). Including the TESS data, our best-fitting orbital decay model gives a period change rate of $\dot{P} = -8.77 \pm 0.47$ ms/yr (Figure A11). Our LOOCV analysis finds that the removal of one data point from literature ($2457824.94673 \pm 0.000009$ BJD$_{\text{TDB}}$, Stefansson et al. 2017) results in the reversal of the sign of the period change rate, with a best-fitting value of $4.38 \pm 0.76$ ms/yr (Figure A12). Subsequently, further analysis of the BIC diagram showes that the removal of another point ($2456885.79665 \pm 0.00008$ BJD$_{\text{TDB}}$, Saeed et al. 2020) also produces a significant decrease in BIC$_{quad}$. Given that these two data points were obtained via ground-based telescopes and possess notably lower uncertainty values compared to the remaining data, it is plausible that the accuracy of these two data points has been overestimated. After the exclusion of both data points, our model fitting yields a period change rate of $-2.74 \pm 0.89$ ms/yr with $\Delta BIC = 9.82$. Thus, we consider TrES-3 b as a mediocre candidate with a weak orbital decay trend.

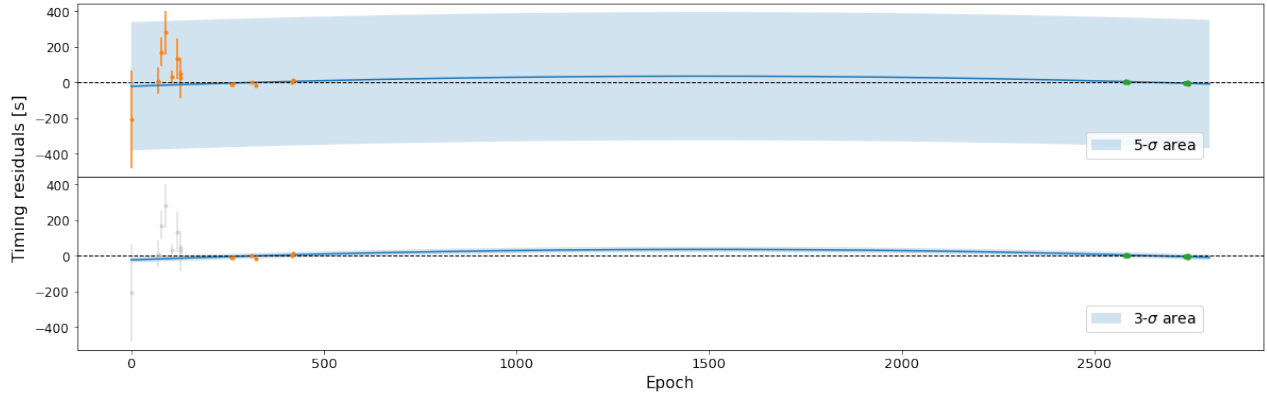

**Figure 6.** Timing residuals of HD 189733 b. The lines and symbols used are similar to Figure 4.

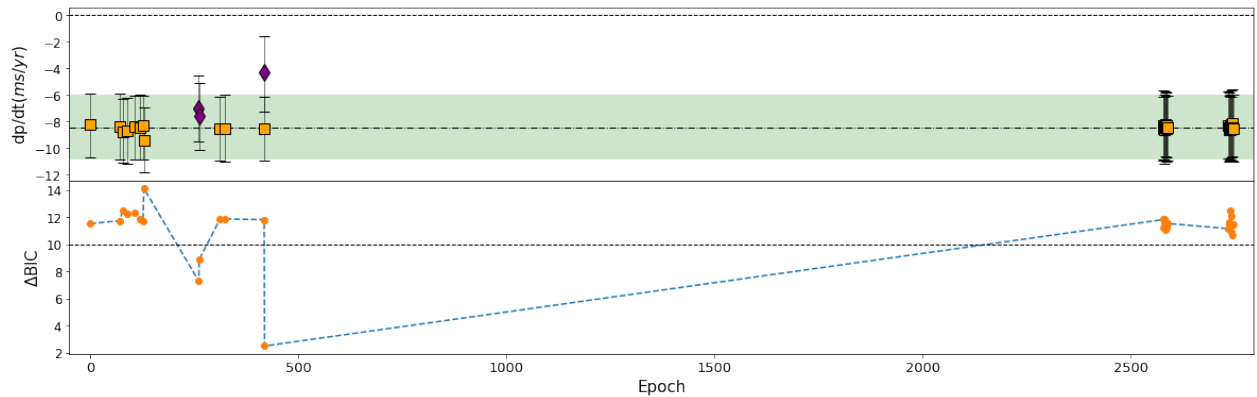

**Figure 7.** LOOCV analysis and corresponding ΔBIC of HD 189733 b. The top panel displays the period change rate dp/dt obtained by fitting the quadratic model after the removal of each single transit timing data. The orange squares show the dp/dt values that satisfy the criterion of being 3σ away from zero, while the purple diamonds represent dp/dt values that fail to meet this criterion. The dash-dotted line marks the original best-fitting dp/dt value before the removal of any data, and the green shaded area marks the corresponding 1σ confidence region. The bottom panel displays the corresponding ΔBIC, where it is evident that the three data points failing the 3σ test also do not satisfy the criterion of ΔBIC< 10. The complete mosaic figure set is available in the online version of this Paper.

### 4.1.14. *WASP-19 b*

WASP-19 b is an ultra-short-period planet first identified by Hebb et al. (2010), with a mass of 1.1 $M_J$ and radius of 1.4 $R_J$. It orbits a G8V star on a 0.789-day period. It is one of the most favorable targets in the search for tidal orbital decay. Patra et al. (2020) found strong evidence of decay with a period change rate of $-6.50 \pm 1.33$ ms/yr. However, the study of Petrucci et al. (2020) showed evidence favors a constant period model and provided an upper limit of $\dot{P}$ as $-2.294$ ms/yr. The latest quadratic analysis performed by Ivshina & Winn (2022) yielded $-3.54 \pm 1.18$ ms/yr. We have analyzed 142 mid-transit time data (87 from literature, 55 from TESS). Our best-fitting decay rate is $\dot{P} = -1.64 \pm 0.29$ ms/yr with two additional sectors of TESS data (Figure A13). In the LOOCV analysis we find that the removal of one data point near 2457796.59224 JD from Espinoza et al. (2019) resulting in a diminishing decay rate of $-0.88 \pm 0.43$ ms/yr. Similar to the case of TrES-3 b, when we remove one additional data point near 2457448.71292 JD from Petrucci et al. (2020) that can significantly reduces the BIC in our LOOCV test, the new rate of orbital decay is $-1.40 \pm 0.30$ ms/yr (Figure A14).

### 4.1.15. XO-3 b

XO-3 b is a 11.7 $M_J$, 1.2 $R_J$ hot Jupiter discovered by Johns-Krull et al. (2008). It orbits an F5V star in a 3.192-day period, with an eccentric and misaligned orbit (Hébrard et al. 2008). TESS transit timing analysis of this system was performed by Yang & Wei (2022), who found $dP/dE = -6.2 \times 10^{-9} \pm 2.9 \times 10^{-10}$, which is equivalent to $\dot{P} = -195 \pm 9$ ms/yr. Subsequent study of Ivshina & Winn (2022) found a decay rate of $-182.08 \pm 12.96$ ms/yr. We have analyzed 47 mid-transit time data (35 from literature, 12 from TESS). With the inclusion of one new sector of TESS data, our best-fitting result is $\dot{P} = -31.59 \pm 5.24$ ms/yr using the 5-$\sigma$ rejection scheme and $-27.67 \pm 5.21$ ms/yr using the 3-$\sigma$ rejection scheme (Figure A15). Through our LOOCV analysis, we find that the primary contributor to the decay trend is the transit time near 2456419.0441 JD derived by Wong et al. (2014) using Spitzer IRAC 4.5 $\mu m$ band data. The exclusion of this point would reduce the decay rate to a much smaller value of $-8.83 \pm 5.14$ ms/yr and a new $\Delta$BIC=2.38 (Figure A16). Thus we do not consider XO-3 b as a strong orbital decay candidate.

### 4.1.16. WASP-16 b

WASP-16 b is a 0.9 $M_J$, 1.0 $R_J$ planet in a 3.12-day orbit around a G3V star (Lister et al. 2009). We have analyzed 17 mid-transit time data (11 from literature, 6 from TESS). Including the TESS data, our best-fitting model gives $\dot{P} = -51.98 \pm 12.87$ ms/yr (Figure A17). In the process of LOOCV analysis, we observed that the removal of one data point (2456037.70089 $\pm$ 0.00024 BJD_TDB, derived from TRESCA by Southworth et al. (2013)) led to the disappearance of the period decay trend (Figure A18). Thus, we consider WASP-16 b as a mediocre candidate.

### 4.1.17. WASP-47 b

WASP-47 b is a 1.1 $M_J$, 1.1 $R_J$ planet in a 4.159-day orbit around a G9V star (Hellier et al. 2012). The WASP-47 system has attract attention due to its remarkable four-planet configuration. WASP-47 b has two nearby neighbors: an interior super-Earth with an orbital period of 0.79 day (WASP-47 e) and an exterior hot-Neptune with an orbital period of 9.0 days (WASP-

47 d). The system also has a distant, moderately eccentric gas giant (WASP-47 c). We have analyzed 31 mid-transit time data (27 from literature, 4 from TESS). Our model fitting yields a decay rate of $\dot{P} = -48.45 \pm 14.14$ ms/yr (Figure A19). The analysis is complicated by the short-term TTV perturbations produced by the other planets in the system. The LOOCV analysis suggests that the long-term trend may not be true, as the removal of the first data point results in the complete disappearance of the trend (Figure A20).

### 4.1.18. WASP-80 b

WASP-80 b is a 0.5 $M_J$, 1.0 $R_J$ planet in a 3.068-day orbit around a K7V star (Triaud et al. 2013). We have analyzed 33 mid-transit time data (28 from literature, 5 from TESS). Our model fitting finds a decay rate of $\dot{P} = -23.04 \pm 5.15$ ms/yr using the 5$\sigma$ rejection scheme and $\dot{P} = -31.99 \pm 4.91$ ms/yr using the 3$\sigma$ rejection scheme. The discrepancy arises from the inclusion of one data point near 2456459.80958 JD, which appears to be an outlier (Figure A21). Our LOOCV analysis identified the transit time data near 2456125.42 JD from Triaud et al. (2013) as the primary contributor to the observed period decay trend (Figure A22). As this data point was collected using a ground-based telescope, it is possible that its precision had been overestimated. Hence, the current data do not provide strong support for WASP-80 b as having significant orbital decay.

## 4.2. Candidates with Increasing Orbital Periods

We have 8 hot Jupiters showing signs of orbital period increase. Among them, 5 systems (4.2.4-4.2.8) have failed to pass the LOOCV test, i.e. upon the removal of one single data point, they no longer meet the two criteria we set above for identifying long-term orbital period variation candidates.

### 4.2.1. HAT-P-7 b

HAT-P-7 b is an ultra-hot Jupiter discovered by Pál et al. (2008), with a mass of 1.8 $M_J$ and radius of 1.5 $R_J$. It orbits an F8 star in a 2.205-day period, likely possessing a retrograde or near-polar orbit with an inclination angle of $\phi \approx 120°$ (Campante et al. 2016; Benomar et al. 2014). Additionally, a common proper motion stellar companion has been identified within this system through high contrast imaging technique (Narita et al. 2012; Ngo et al. 2015). This planet has received significant attention due to its unusual atmospheric and orbital properties. We have analyzed 77 mid-transit time data (11 from literature, 66 from TESS). Our analysis of long-term archival and TESS transit time data indicates the planet has a positive period change rate of $\dot{P} = 18.28 \pm 4.05$ ms/yr (Figure A23).

### 4.2.2. WASP-1 b

WASP-1 b is a 0.9 $M_J$, 1.5 $R_J$ planet in a 2.52-day orbit around a F7V star (Collier Cameron et al. 2007). A common proper motion stellar companion has been identified within this system (Ngo et al. 2015; Collier Cameron et al. 2007) . We have analyzed 45 mid-transit time data (36 from literature, 9 from TESS). Our quadratic model fitting yields a best-fitting period change rate of $\dot{P} = 22.50 \pm 5.23$ ms/yr using $5\sigma$ rejection scheme and $24.33 \pm 5.19$ ms/yr using $3\sigma$ rejection scheme, respectively (Figure A24).

### 4.2.3. WASP-46 b

WASP-46 b is a 1.9 $M_J$, 1.2 $R_J$ hot Jupiter discovered by Anderson et al. (2012). It orbits a G6V star in a 1.430-day period. The host star exhibits weak Ca II H+K emission, indicating an active photosphere and chromosphere. Ciceri et al. (2016) conducted the first TTV study of this planet, finding that a linear ephemeris model yields an inadequate fit to the observations. Petrucci et al. (2018) investigated homogeneous TTVs for the planet using both new and previously published data, concluding that the potential for orbital decay cannot be excluded. Davoudi et al. (2021) observed a positive orbital variation and attributed it to star magnetic activity. We have analyzed 86 mid-transit time data (54 from literature, 32 from TESS). Our analysis yielded an increasing rate of $\dot{P} = 51.68 \pm 2.86$ ms/yr using the $5\sigma$ rejection scheme and $63.22 \pm 2.78$ ms/yr using the $3\sigma$ rejection scheme, respectively (Figure A25). However, the large scattering shown in the residuals casts doubts on the ability of the quadratic model to explain the observations.

### 4.2.4. HAT-P-43 b

HAT-P-43 b is a 0.7 $M_J$, 1.3 $R_J$ planet in a 3.333-day orbit around a G-type star (Boisse et al. 2013). We have analyzed 48 mid-transit time data (7 from literature, 41 from TESS). Our best-fitting model shows a period change rate of $\dot{P} = 94.40 \pm 22.55$ ms/yr using the $5\sigma$ rejection scheme and $108.33 \pm 23.91$ ms/yr using the $3\sigma$ rejection scheme, respectively (Figure A26). However, upon removal of the first data point adopted from Boisse et al. (2013) in our LOOCV test, we find the system no longer meets the criterion to be classified as long-term orbital period variation candidate (Figure A27). It is therefore a mediocre candidate for long-term period variation, and further observations are necessary to help clarify this result.

### 4.2.5. HAT-P-44 b

HAT-P-44 b is a 0.35 $M_J$, 1.2 $R_J$ hot Jupiter discovered by Hartman et al. (2014). It orbits a G8V star in a 4.301-day period. An outer planet, HAT-P-44 c, with a mass of at least 1.6 $M_J$ and a period of 219.9 days, is likely present in the system (Hartman et al. 2014). We have analyzed 27 mid-transit time data (13 from literature, 14 from TESS). Our analysis of HAT-P-44 b's TTV has unveiled an upward trend in the orbital period variation at a rate of $\dot{P} = 119.67 \pm 25.89$ ms/yr (Figure A28). In our LOOCV analysis, the removal of the first data point near 2455696.94 JD from Hartman et al. (2014) results in a new period change rate of $144.47 \pm 57.56$ ms/yr, which falls short of the $3\sigma$ requirement needed to be identified as an orbital period variation candidate (Figure A29). Nevertheless, this trend does not disappear entirely and further observation would be necessary to confirm its existence.

### 4.2.6. WASP-6 b

WASP-6 b is a 0.5 $M_J$, 1.2 $R_J$ planet in a 3.361-day orbit around a G8 star (Gillon et al. 2009a). We have analyzed 25 mid-transit time data (12 from literature, 13 from TESS). Our best-fitting model shows a period increase rate at $\dot{P} = 20.66 \pm 5.19$ s/yr (Figure A30). Our LOOCV analysis indicated that once the first data point near 2454596.43 JD adopted from Gillon et al. (2009a) was removed, the period increase rate would be reduced to $13.87 \pm 6.22$ ms/yr, which fails to meet the $3\sigma$ criterion for orbital variation (Figure A31). Nevertheless, this positive period change rate still deviates more than $2-\sigma$ away from zero, and additional data are warranted to refute the existence of positive $\dot{P}$.

### 4.2.7. WASP-11 b

WASP-11 b (HAT-P-10) is a 0.5 $M_J$, 1.0 $R_J$ hot Jupiter discovered by West et al. (2009) and Bakos et al. (2009). It orbits around a K3V star in a 3.722-day period. The presence of a low-mass stellar companion in the system has been confirmed by both direct imaging and a RV trend (Knutson et al. 2014; Ngo et al. 2015). We have analyzed 35 mid-transit time data (24 from literature, 11 from TESS). Our analysis of its orbital period variation yields an increasing $\dot{P}$ of $23.32 \pm 6.69$ ms/yr (Figure A32). In our LOOCV analysis, the exclusion of one data point near 2455898.76 JD adopted from Wang et al. (2014) would reduce the $\dot{P}$ to $17.13 \pm 6.63$ ms/yr (Figure A33). This removed transit time data point from ETD was originally derived using self-reported data from a ground-based telescope (Wang et al. 2014), which may not be very reliable. The new derived $\dot{P}$ still sits near the $3\sigma$ border, which renders it as a good candidate.

### 4.2.8. WASP-17 b

WASP-17 b is a 0.5 $M_J$, 1.9 $R_J$ planet in a 3.735-day orbit around a F4 star (Anderson et al. 2010). We have analyzed 31 mid-transit time data (19 from literature, 12 from TESS). Our best-fitting model has a positive period change rate of $\dot{P} = 77.64 \pm 8.19$ ms/yr (Figure A34). Our LOOCV analysis has revealed that the transit data near 2454559.19 JD derived by Bento et al. (2014) using ULTRACAM on ESO's NTT is the most significant contributor to the observed increasing trend. It is likely that the reported uncertainty of this transit time data point has been underestimated. After the removal of this data point, the re-fitted model shows a trend of $3.36 \pm 4.46$ ms/yr, which is consistent with a constant orbital period (Figure A35). Thus, we consider WASP-17 b as a mediocre candidate.

### 4.3. *Constant Period Candidates*

Analyses of transit time data of the remaining 300 hot Jupiters reveal no strong evidence of orbital period change. 133 of them show a good fit by both the linear and quadratic models but do not meet the $3 - \sigma$ criterion, while the other 167 hot Jupiters shows only a good fit by the linear model. We have shown one such example, HAT-P-16 b, in Figure 8. We can use the quadratic model to fit the data well with a best-fit period change rate value of $\dot{P} = 3.41 \pm 4.99$ ms/yr, which is consistent with a zero value period change rate. To assess the period change rate measurement precision, we plot the distribution of the 3-sigma error bars of the period change rate $\dot{P}$ for the 133 hot Jupiters with a good quadratic model fit. (Figure 9) As can be seen from this distribution, we can detect an absolute period change rate $|\dot{P}|$ greater than $\sim 100$ ms/yr for most of these hot Jupiters. The lack of enough transit time data is the main cause for the other 167 hot Jupiters to have a bad quadratic fit. More high precision transit time data in the future are needed to put a more tight constraint on their period change rate, such as from PLAnetary Transits and Oscillation of stars mission and the China Space Station Telescope (PLATO, expected around 2026, Rauer et al. 2014), (CSST, expected around 2025, Zhan 2011) and the Earth 2.0 mission (ET, expected around 2027, Ge et al. 2022).

## 5. DISCUSSION

### 5.1. *LOOCV Test*

The LOOCV test is an effective method of highlighting the significant impact of a single deviant mid-time measurement (or an underestimated error) on the corresponding $\Delta BIC$ diagram. There are a total of 7 out of 18 of systems having high $\Delta BIC$ values that result from a single data point for period decrease candidates, and 5 out of 8 for period increase candidates, which are summarized in Table 3. This indicates that $\Delta BIC$ is fairly sensitive even to the presence of a single outlier data point.

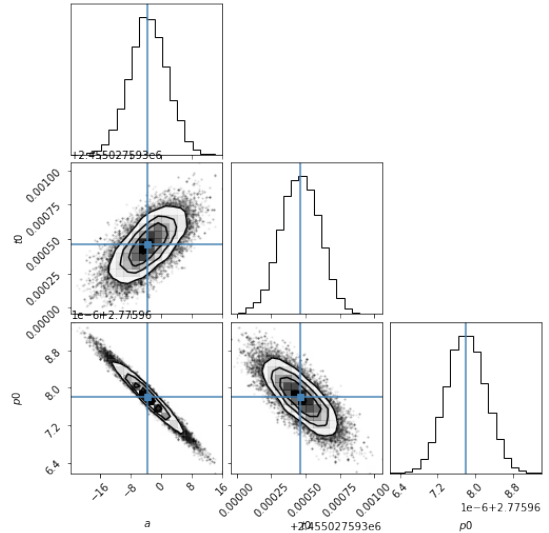

**Figure 8.** Posterior distribution of the fitted model parameters of HAT-P-16 b from our MCMC quadratic model fitting.

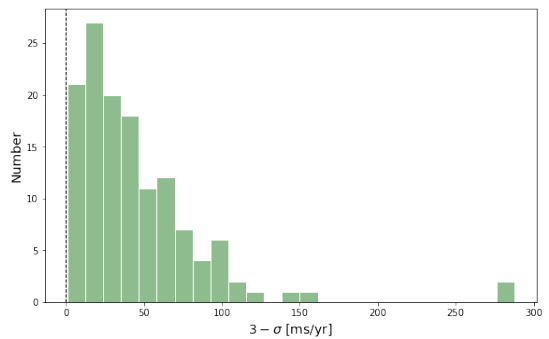

**Figure 9.** Distribution of the 3-sigma error bars of the period change rate $a$ for the 133 hot Jupiters with a good quadratic model fit. As can be seen from this distribution, we can detect an absolute period change rate $|a|$ greater than $\sim 100$ ms/yr for most of these hot Jupiters.

The problem mid-transit time points in the system of WASP-47 b, HAT-P-43 b, HAT-P-44 b, and WASP-6 b are all composite times, which have been derived by simultaneously fitting photometry data and radial velocity data obtained from different instruments spanning one to several years. Thus these data points could have larger than expected systematic noises when comparing with mid-transit time data derived by fitting individual transits and should be used cautiously when investigating the TTV of individual systems. For WASP-16 b, WASP-11 b, XO-3 b and HD 189733 b, the problem points are likely caused by unknown systematic er-rors when de-trending the photometry light curves. For TrES-3 b, WASP-19 b, WASP-80 b and WASP-17 b, the problem points likely have unreasonably small error bars than expected from similar ground-based observations. These issues serve as a reminder that prudence should be exercised when employing data from the literature.

## 5.2. Comparison to Previous Works

Several groups have published long-term transit time variation study en masse (Patra et al. 2020; Ivshina & Winn 2022; Shan et al. 2023). We compare our results with their results in this section.

### 5.2.1. Comparison to Ivshina & Winn (2022)

Ivshina & Winn (2022) have compiled a database of transit times for 382 planets using data from the TESS mission and previously published transit times from literature, and use the database to update ephemerides . They also use the database to discover 10 cases with suggestive evidences for orbital period changes. We have used part of the data from their database, it is important for us to make a detailed comparison here.

For the 10 planet candidates flagged as exhibiting long-term period changes, we first use our code to fit their transit time data and find similar fitting results of $\dot{P}$ as shown in Ivshina & Winn (2022), usually within the $1\sigma$ error bar. This proves the robustness of our code, and the main discrepancy between results from this work and their work is the transit time data used, where we have additional TESS data and a longer time span for the ephemeris model fitting. For example, all but the three planets, WASP-4 b, WASP-45 b, and WASP-99 b, have new TESS data available in this work.

For WASP-12 b, WASP-4 b, TrES-1 b, WASP-99 b and WASP-45 b, consistent results of the period change rate (within $1\sigma$) have been obtained. For TrES-5 b, WASP-19 b, and XO-3 b, a smaller absolute period change rate ($|\dot{P}|$) are obtained with the updated TESS data from this work. For CoRoT-2 b, we have made adjustments to the apparently unreasonable error bars adopted in some of the literature data and obtained a significantly smaller absolute period change rate. For WASP-161 b, we will discuss it in the next section, Section 5.2.2.

There are 8 hot Jupiters flagged as period decay cases in this work but not flagged by Ivshina & Winn (2022). For HAT-P-37 b, WASP-80 b and KELT-16 b, the additional TESS observations available at the time of this study have caused the fitting difference. However, for HD 189733 b, WASP-16 b, WASP-22 b, and WASP-47 b, we can still derive period change rates that meet the 3-sigma criteria using only data from Ivshina & Winn (2022), which suggests it is likely they have masked some of the transit time data during their fitting process. Another special case is TrES-3 b. Three middle transit time data points adopted from literature by Ivshina & Winn (2022) have unrealistic small error bars (as small as 1 second), where two of them showing very big impact on the model fitting evidenced by our LOOCV analysis. Ivshina & Winn (2022) have taken care of only one data point by manually inflating its error bar by a factor of 30, while we manually remove both data points. This is the main reason why this planet has failed to meet the selection criteria for being a period change candidate in Ivshina & Winn (2022) but meet the selection criteria in this work. When we manually inflate the error bars of all these three data points to a value of about 30 second, we can still obtain a negative period change rate.

Ivshina & Winn (2022) did not report any candidate having an increasing period. For HAT-P-7 b, HAT-P-43 b, HAT-P-44 b, WASP-1 b, and WASP-11 b, the additional TESS observations available at the time of this study have made the fitting difference. On the other hand, for WASP-6 b, WASP-17 b, and WASP-46 b, we can still identify them as period increasing candidates when fitting the data from Ivshina & Winn (2022) only. This could be due to the fact that Ivshina & Winn (2022) were not interested in finding candidates with a positive period change rate.

### 5.2.2. Comparison to Shan et al. (2023)

Shan et al. (2023) have analyzed the transit timing data from the TESS Objects of Interest Catalog for a sample of 262 hot Jupiters. They identified 31 hot Jupiters for which the TESS mid transit times show at least $1\sigma$ offsets from predictions calculated using ephemerides taken from the literature. They used tidal dissipation to explain some of the timing offsets seen in their results. Our approach differs from theirs in that we do not rely on previously published ephemerides but re-do a new ephemeris model fitting using the new TESS data and archival data. We then use the new ephemeris model fitting results to identify possible candidates showing a positive or negative long-term orbital period change rate.

XO-3 b and WASP-17 b are the only two systems on both of our interested candidate lists. For XO-3 b, as we have already pointed out in section 4.1.15, the earlier work of Yang & Wei (2022) reported a decay rate of $-195 \pm 9$ ms/yr, whereas in this work we have obtained a much smaller value of $-32 \pm 5$ ms/yr with additional TESS data. For WASP-17 b, Shan et al. (2023) found it shows the largest late offset timing of $70.8 \pm 11.7$ minutes using a linear ephemeris model, while we find it to be consistent with $\dot{P} = 0$ ms/yr after the removal of one single transit time data with a very small error bar ($\sim$ 8 second).

Another target worth mentioning on their candidate list is WASP-161 b. This is another case that demonstrates the importance of acquiring long-term high quality transit time data in studying the long-term orbital period variation of hot Jupiters. Shan et al. (2023) suggested that WASP-161 b is undergoing tidal dissipation with $\dot{P} = -1.16 \times 10^{-7} \pm 2.25 \times 10^{-8}$, corresponding to $\dot{P} = -3.66 \pm 0.7$ s/yr. Ivshina & Winn (2022) reported a $\dot{P}$ result of $-15.5 \pm 0.35$ s/yr and suggested that there may be an error when obtaining the transit time data by Barkaoui et al. (2019). Yang & Chary (2022) have re-analyzed the SSO-Europa light curve data used by Barkaoui et al. (2019) and obtained a new mid transit time of 2458124.71742±0.00083 HJD, which can be converted to 2458124.718220742±0.00083 BJD$_{\mathrm{TDB}}$. We here adopt this new transit time from Yang & Chary (2022) instead of the time data from Barkaoui et al. (2019) in our model fitting. Combining with new TESS data available in 2023 January, we have derived a new orbital period change rate of $\dot{P} = -0.786 \pm 0.208$ s/yr, which is much lower than the results reported by Ivshina & Winn (2022) and Shan et al. (2023). Further observations with precise timing are needed to better understand its orbital period variation trend.

### 5.2.3. *Comparison to Patra et al. (2020)*

In the study of Patra et al. (2020), they have analyzed long-term transit timing data for 12 hot Jupiters orbiting bright host stars to seek direct evidence of tidal orbital decay. Three of their targets are overlapping with our long-term period change candidates (WASP-12 b, KELT-16 b and WASP-19 b), which have been discussed in section 4.1.1, 4.1.5, and 4.1.14. We here discuss another five targets from their sample.

WASP-18 b is a 10.4 $M_J$, 1.2 $R_J$ "super-Jupiter" in a 0.941-day orbit around a F6 star (Hellier et al. 2009). It was considered as a TTV candidate and has been extensively studied (Wilkins et al. 2017; McDonald & Kerins 2018; Shporer et al. 2019; Patra et al. 2020; Maciejewski et al. 2022; Rosário et al. 2022). They do not find any sign of orbital decay in the system. Our best-fit result of the period change rate is $\dot{P} = 0.73 \pm 0.75$ ms/yr, which is consistent with a constant period model.

HATS-18 b is a 2.0 $M_J$, 1.3 $R_J$ planet in a 0.838-day orbit around a G-type star (Penev et al. 2016). It has received a lot of attention as an orbital decay candidate. However, all the recent studies from Penev et al. (2016), Patra et al. (2020), and Southworth et al. (2022) do not find strong evidence of periodic variation in HATS-18 b.

Our analysis in this work yields a period change rate of $-6.72 \pm 3.01$ ms/yr and $\Delta BIC = 5$, which slightly favors a quadratic model than a linear model.

HAT-P-23 b is a 2.1 $M_J$, 1.2 $R_J$ planet in a 1.213-day orbit around a G5 star (Bakos et al. 2011). Maciejewski et al. (2018), Patra et al. (2020) and Baştürk et al. (2022) have studied its transit timing data and found it to be consistent with a constant period model. Our quadratic model fitting yields a period change rate of $1.46 \pm 1.52$ ms/yr, which is also in good agreement with a constant period model.

WASP-43 b is a 2.0 $M_J$, 1.0 $R_J$ planet in a 0.813-day orbit around a K7V star (Hellier et al. 2011b). It has been the subject of discussion as an orbital decay candidate for a long period of time. Ever since the first report of orbital decay from WASP-43 b by Jiang et al. (2016), all subsequent studies have effectively ruled out the existence of orbital decay by incorporating new timing data (Hoyer et al. 2016; Stevenson et al. 2017; Patra et al. 2020; Garai et al. 2021; Davoudi et al. 2021; Hagey et al. 2022). Our quadratic model fitting yields a period change rate of $0.19 \pm 0.34$ ms/yr, which also supports the constant period scenario. WASP-43 b serves as another notable example that illustrates more high quality timing data can lead to a more accurate result.

WASP-122 b is a 1.4 $M_J$, 1.8 $R_J$ planet in a 1.710-day orbit around a G4 star (Turner et al. 2016b). Due to its favorable physical properties, Patra et al. (2020) have put this system on their list as a target to watch for orbital decay. We obtain a best-fit period change rate of $0.18 \pm 4.25$ ms/yr using archival timing data from literature and four sectors of TESS data for WASP-122 b. This result agrees well with a constant period model, showing no evidence of orbital decay.

## 5.3. *Possible Explanation for Long-term Period Variation*

There are several possible scenarios to explain the long-term orbital period variations of exoplanets. Here we summarize a few of them in this section, including Rømer effect, apsidal precession, the Applegate effect, mass loss and tidal dissipation.

**Rømer effect**. Due to variations in light travel time, the accelerating motion of the center of mass of the system towards or away from the observer's line of sight can cause illusory changes in the period, which is known as the Rømer effect (Bouma et al. 2019). This acceleration is usually attributed to a wide-orbit companion. Acceleration towards the observer results in a decreasing period, while an increasing period corresponds to acceleration away from the observer. In fact, several of our candidate systems exhibit evidence of companions on a wide orbit, as discussed in Section 4. Because the observed period derivative of hot Jupiters would be correlated with the time derivative of the RV data, as shown in Equation 23 of Bouma et al. (2019), the Rømer effect can be verified through the fitting of RV data of the host star.

**Apsidal precession**. Apsidal precession is expected in systems where hot Jupiters are in orbits that are at least slightly eccentric ($e > 0.003$), causing sinusoidal variations in transit times. It's driven mainly by the quadrupole of the planetary tidal bulge, with additional contributions from general relativity (Ragozzine & Wolf 2009). For the most promising systems, the precession period can be as short as a few decades (Birkby et al. 2014). Therefore, within a relatively short observational window, the curvature of TTVs from Apsidal precession can resemble signals of orbital decay or expansion. Apsidal precesssion has been ruled out for some systems (WASP-12b, e.g., Patra et al. 2017; Yee et al. 2020; Turner et al. 2021). This could be due to the fact that hot Jupiters tend to have very low eccentricities. Unless triggered by some special conditions such as an external perturber, tidal interactions tend to circularize the orbits of hot Jupiter on a very short timescale. For most of our candidates, the higher limits of their measured eccentricities are well above 0.003. Thus we cannot rule out the possibility that Apsidal precession may account for some of period variations derived in this work. We anticipate the forthcoming transit and occulation observations in the next 5-10 years can help reveal the nature of apsidal precession from hot Jupiters.

**The Applegate effect.** The Applegate effect (Applegate 1992) offers an explanation for quasi-periodic variations over times-scales of years to decades in eclipse times of eclipsing binaries. This effect suggests that magnetically active stars undergo cyclic changes in their internal structure, resulting in the exchange of angular momentum between their inner and outer zones. While originally applied to eclipsing binaries, this mechanism could also apply to hot Jupiters orbiting a star with a convective zone (Watson & Marsh 2010). The changing gravitational quadrupole of the star would cause the planet's orbital period to vary over the timescale of the stellar activity cycle. The magnitude of TTV driven by the Applegate mechanism is proportional to $a^{-2}T_{\mathrm{modulation}}^{3/2}$ (Watson & Marsh 2010), which is particular strong for hot Jupiter due to its small star-planet separation. The modulation time-scale $T_{\mathrm{modulation}}$ is related to the period of stellar activity cycle, which could be equal to 11 or 22 yr for our Sun as an example depending on the dynamo mechanism at work. Watson & Marsh (2010) have estimated that TTV amplitudes caused by the Applegate effect range from a few seconds for modulation time-scale of 11 yrs to several minutes for modulation time-scale of 50 yrs. For short period transiting hot Jupiters, such as WASP-18 b, the time-scales and amplitudes of Applegate driven TTVs are comparable to those of TTVs induced by tidal dissipation (Watson & Marsh 2010). Thus, The Applegate effect may account for part of the TTVs observed in some of our candidates over decades long time-scales, and further investigation is needed to verify its presence.

**Mass loss.** When a hot Jupiter approaches its Roche limit, mass loss could occur, which may be due to escaping winds or Roche lobe overflow. This mass loss leads to a decrease in planetary density, causing the Roche limit to expand outward. Consequently, the torques in the accretion disk act to exchange angular momentum with the planet. These torques can drive the planet away from the star, causing it to migrate outward with the expanding Roche limit. The subsequent evolutionary scenarios depend on various characteristics of the system, such as the core mass of the planet. Whether mass loss ultimately leads to orbital expansion or contraction depends on various conditions (Valsecchi et al. 2015; Jackson et al. 2016). Several of our samples are close to the Roche limit or even expected to undergo mass loss at extremely high rates, such as WASP-12 b and WASP-121 b (Li et al. 2010; Salz et al. 2019). Although the mass loss effect may not explain the period increase seen in our candidates, it remains an intriguing mechanism warrants further investigation.

**Tidal effects.** As is known, in the case where the orbital period of a hot Jupiter is shorter than the rotation period of its host star, the hot Jupiter is prone to undergo inward spiral migration (Rasio et al. 1996; Levrard et al. 2009; Matsumura et al. 2010). Conversely, when the rotation period of the host star is shorter than the orbital period of the hot Jupiter, there is a possibility of angular momentum transfer from the rapidly rotating host star to the planet's orbit, leading to an increase in the orbital period (Ogilvie 2014). However, for our orbital period increasing candidates in Section 4.2, both the stellar rotation periods derived from the literature and the ones we estimated through $v \sin i$ method are longer than the orbital periods of the planets. Therefore, tidal effects are unlikely to be a plausible explanation for orbital period increase.

## 6. CONCLUSION

We have analyzed transit time data from TESS and literature for a total of 326 hot Jupiters, to identify candidates with positive or negative long-term period change rate. We fit the transit time data using both of a linear and a quadratic ephemeris model. We find 18 hot Jupiters showing evidence of negative period change rate and 8 with positive period change rate. Our results are useful to anyone who is interested in planning observations for these systems in the future.

We plan to expand the TTV study of hot Jupiters into the field of short period transiting brown dwarfs (BDs), to help further explore the possible differences between the HJ population and BD population raised by Grether & Lineweaver (2006) and Ma & Ge (2014), etc. For example, Bowler et al. (2020) have argued that high-mass BDs predominantly form like stellar binaries by comparing the mass-eccentricity distribution of BDs to that of giant planets. Since TTV studies can also probe the dynamic properties of the HJs and BDs systems, we expect the TTV studies conducted on both HJ and BD populations can potentially offer more observational evidence supporting distinct formation channels. Thus we encourage our colleagues to continue monitor the transits of not only HJs, but also transiting BDs.

We are indebted to Ivshina & Winn (2022) for the development of the database that served as a valuable resource for our study. Their meticulous efforts in compiling and organizing the data were instrumental in our analysis. We acknowledge the significant contribution of the NASA Exoplanet Archive, which provided access to a wealth of observational data and resources. The availability of such comprehensive databases greatly facilitated our research endeavors.

Furthermore, we extend our appreciation to the TESS mission for its remarkable contribution to exoplanet science. The high-quality transit time measurements obtained from TESS played a crucial role in our investigation of transit timing variations (TTVs) in hot Jupiters.

## APPENDIX

### A. APPENDIX A

Previous literature references for the data used by the 26 candidates in this article:

**WASP-12 b** Hebb et al. (2009) Chan et al. (2011) Sada et al. (2012) Cowan et al. (2012) Copperwheat et al. (2013) Stevenson et al. (2014) Kreidberg et al. (2015) Maciejewski et al. (2016a) Collins et al. (2017) Patra et al. (2017) Patra et al. (2020)

**WASP-4 b** Wilson et al. (2008) Winn et al. (2009) Gillon et al. (2009b) Dragomir et al. (2011) Sanchis-Ojeda et al. (2011) Nikolov et al. (2012) Hoyer et al. (2013) Ranjan et al. (2014) Huitson et al. (2017) Baluev et al. (2019) Southworth et al. (2019) Bouma et al. (2020)

**CoRoT-2 b** Alonso et al. (2008) Vereš et al. (2009) Rauer et al. (2010) Baluev et al. (2015) Öztürk & Erdem (2019)

**HAT-P-37 b** Bakos et al. (2012) Maciejewski et al. (2016b) Turner et al. (2017) Wang et al. (2021) Yang et al. (2021) A-thano et al. (2022)

**KELT-16 b** Oberst et al. (2017) Maciejewski et al. (2018) Patra et al. (2020) Mancini et al. (2022)

**TrES-1 b** Alonso et al. (2004) Charbonneau et al. (2005) Winn et al. (2007a) Narita et al. (2007) Rabus et al. (2009) Hrudková et al. (2009) Cubillos et al. (2014) Wang et al. (2021)

**TrES-5 b** Mandushev et al. (2011) Mislis et al. (2015) Maciejewski et al. (2016b) Sokov et al. (2018) Maciejewski et al. (2021)

**WASP-10 b** Barros et al. (2013) Sada & Ramón-Fox (2016) Zellem et al. (2020)

**WASP-22 b** Maxted et al. (2010) Anderson et al. (2011a) Southworth et al. (2016)

**WASP-45 b** Anderson et al. (2012) Ciceri et al. (2016)

**XO-4 b** Narita et al. (2010) Villanueva et al. (2016)

**HD 189733 b** Winn et al. (2007b) Morvan et al. (2020)

**TrES-3 b** Sozzetti et al. (2009) Gibson et al. (2009) Colón et al. (2010) Lee et al. (2011) Christiansen et al. (2011) Sada et al. (2012) Jiang et al. (2013) Turner et al. (2013) Vaňko et al. (2013) Kundurthy et al. (2013) Püsküllü et al. (2017) Ricci et al. (2017) Stefansson et al. (2017) Sun et al. (2018) Mannaday et al. (2020) Saeed et al. (2020) Saha et al. (2021)

**WASP-19 b** Hebb et al. (2010) Hellier et al. (2011a) Dragomir et al. (2011) Lendl et al. (2013) Tregloan-Reed et al. (2013) Mancini et al. (2013) Petrucci et al. (2020) Patra et al. (2020) Espinoza et al. (2019)

**XO-3 b** Johns-Krull et al. (2008) Winn et al. (2008) Hébrard et al. (2008) Wong et al. (2014) Turner et al. (2017)

**WASP-16 b** Lister et al. (2009) Southworth et al. (2013)

**WASP-47 b** Hellier et al. (2012) Vanderburg et al. (2017) Nascimbeni et al. (2023)

**WASP-80 b** Triaud et al. (2013) Mancini et al. (2014) Fukui et al. (2014) Sedaghati et al. (2017) Turner et al. (2017) Kirk et al. (2018) Wang et al. (2021)

**HAT-P-7 b** Pál et al. (2008) Christiansen et al. (2010) Wong et al. (2016)

**WASP-1 b** Charbonneau et al. (2007) Collier Cameron et al. (2007) Shporer et al. (2007) Szabo et al. (2010) Albrecht et al. (2011) Sada et al. (2012) Granata et al. (2014) Maciejewski et al. (2014) Turner et al. (2016a) Ivshina & Winn (2022)

**WASP-46 b** Anderson et al. (2012) Ciceri et al. (2016) Moyano et al. (2017) Petrucci et al. (2018)

**HAT-P-43 b** Boisse et al. (2013) Mallonn et al. (2019)

**HAT-P-44 b** Hartman et al. (2014) Mallonn et al. (2019) Ivshina & Winn (2022)

**WASP-6 b** Gillon et al. (2009a) Dragomir et al. (2011) Sada et al. (2012) Jordán et al. (2013) Nikolov et al. (2015) Tregloan-Reed et al. (2015)

**WASP-11 b** West et al. (2009) Bakos et al. (2009) Sada et al. (2012) Wang et al. (2014) Ivshina & Winn (2022)

**WASP-17 b** Anderson et al. (2010) Anderson et al. (2011b) Southworth et al. (2012) Bento et al. (2014) Sedaghati et al. (2016)

### B. APPENDIX B

In this section, we show the transit timing data fitting results for the period decay candidates identified in this work, from Figure A1 to A22.

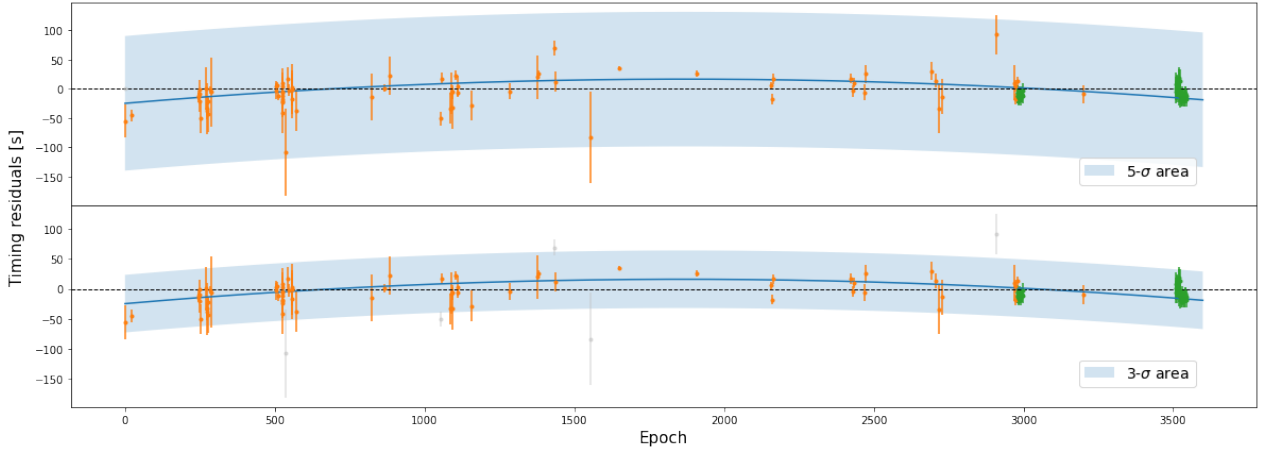

**Figure A1.** Timing residuals of WASP-4 b. The top panel displays the fitting results using a 5-$\sigma$ rejection scheme, while the bottom panel shows the fitting results using a 3-$\sigma$ rejection scheme. The blue curves and shaded areas indicate the best-fit quadratic model and corresponding $5 - \sigma$ (top) and $3 - \sigma$ (bottom) confidence regions. The orange points are based on literature data. The green points are based on TESS data. The gray points are clipped data.

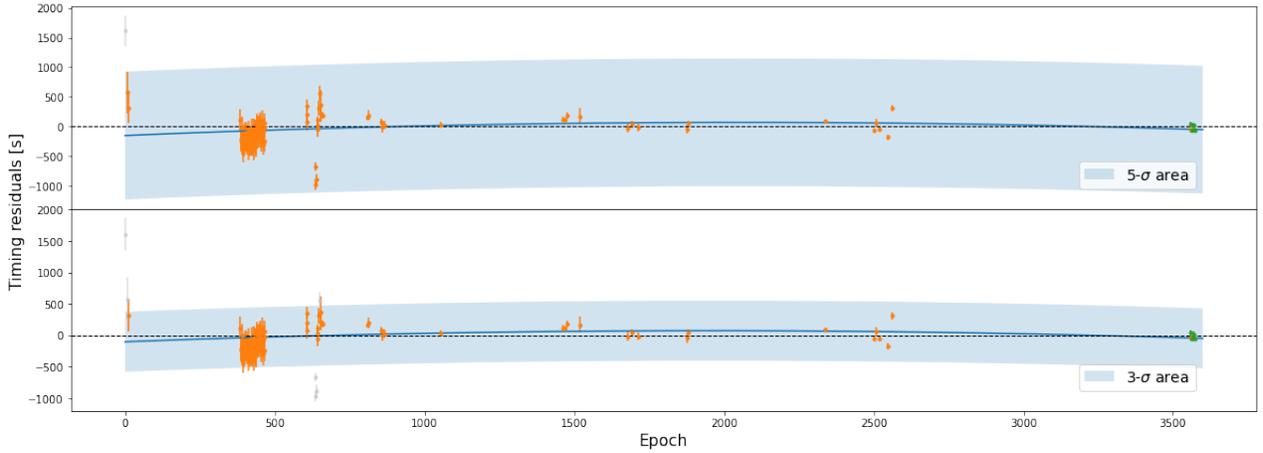

**Figure A2.** Timing residuals of CoRoT-2 b. The lines and symbols used are similar to Figure A1.

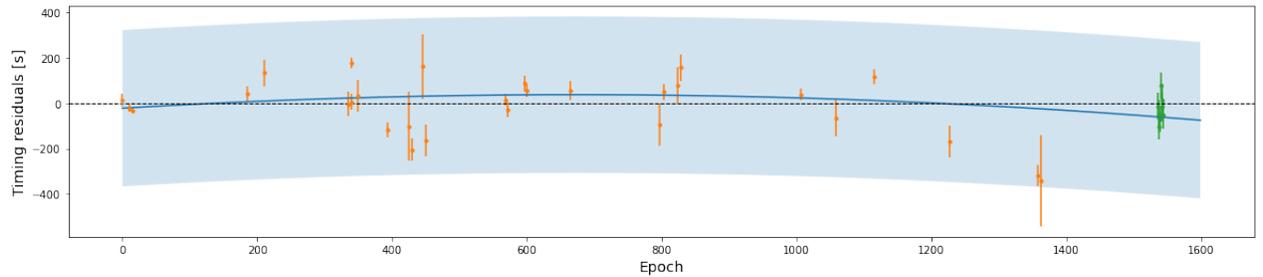

**Figure A3.** Timing residuals of HAT-P-37 b. The lines and symbols used are similar to Figure A1. Since neither 3-sigma rejection nor 5-sigma rejection clipped out any data points, i.e. their results were consistent, we present only the fitting results for 3-sigma rejection.

## C. APPENDIX C

In this section, we show the transit timing data fitting results for the period increasing candidates identified in this work, from Figure A23 to A35.

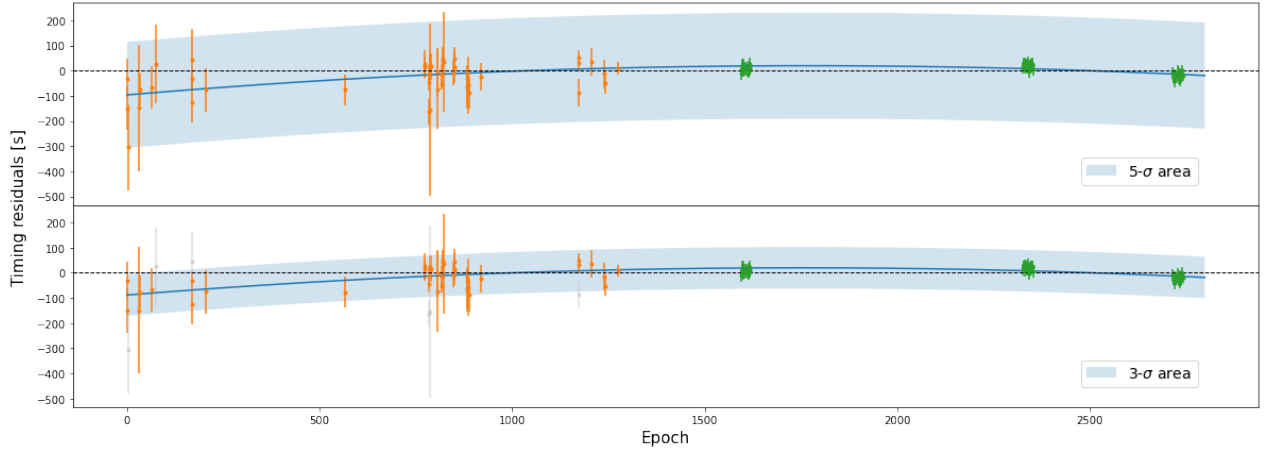

**Figure A4.** Timing residuals of KELT-16 b. The lines and symbols used are similar to Figure A1.

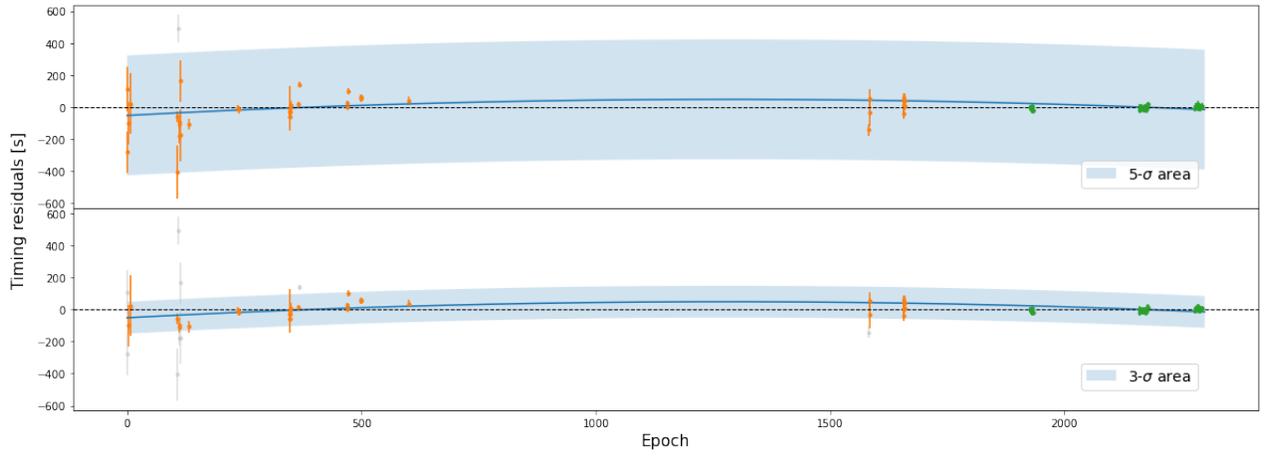

**Figure A5.** Timing residuals of TrES-1 b. The lines and symbols used are similar to Figure A1.

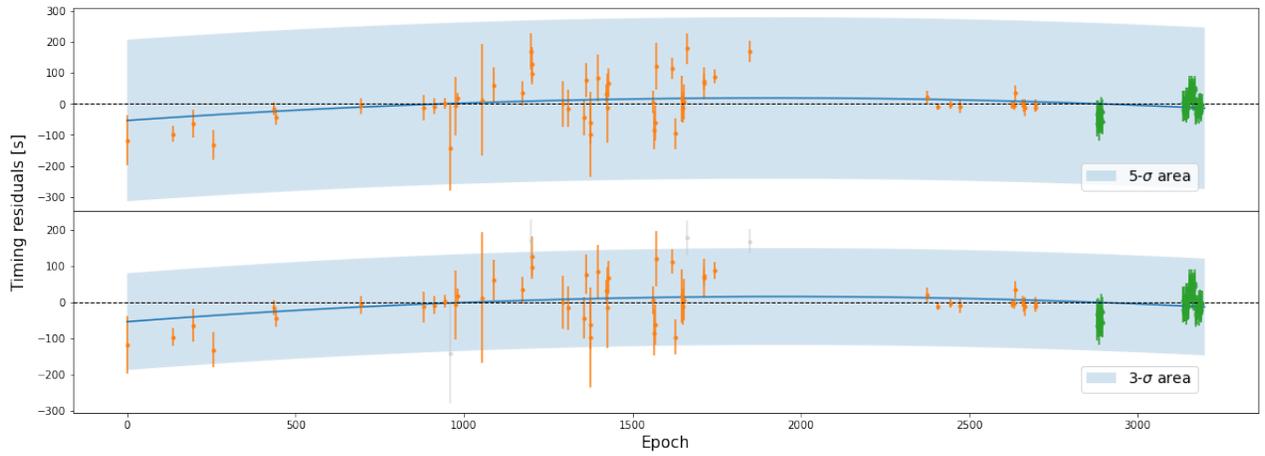

**Figure A6.** Timing residuals of TrES-5 b. The lines and symbols used are similar to Figure A1.

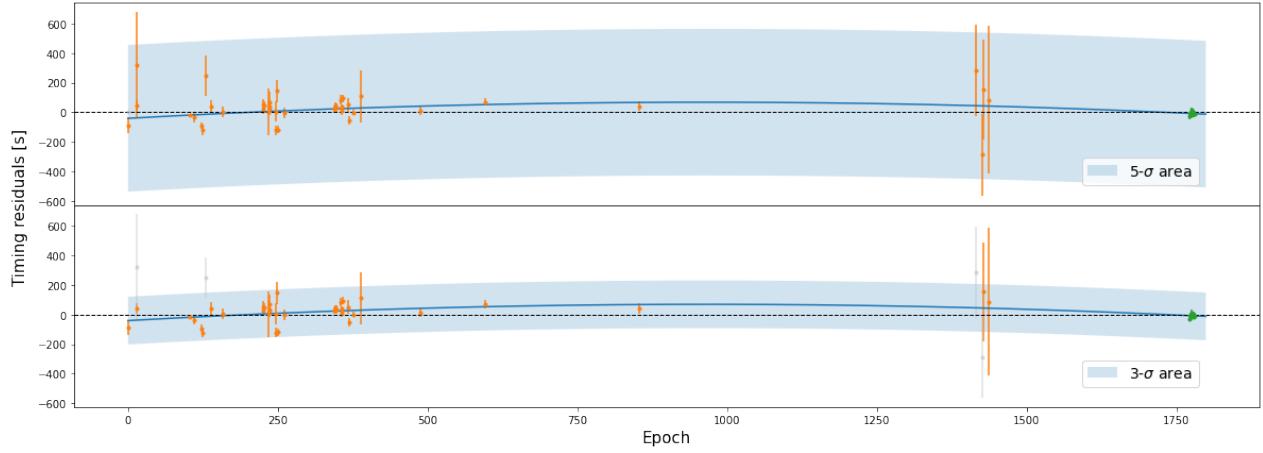

**Figure A7.** Timing residuals of WASP-10 b. The lines and symbols used are similar to Figure A1.

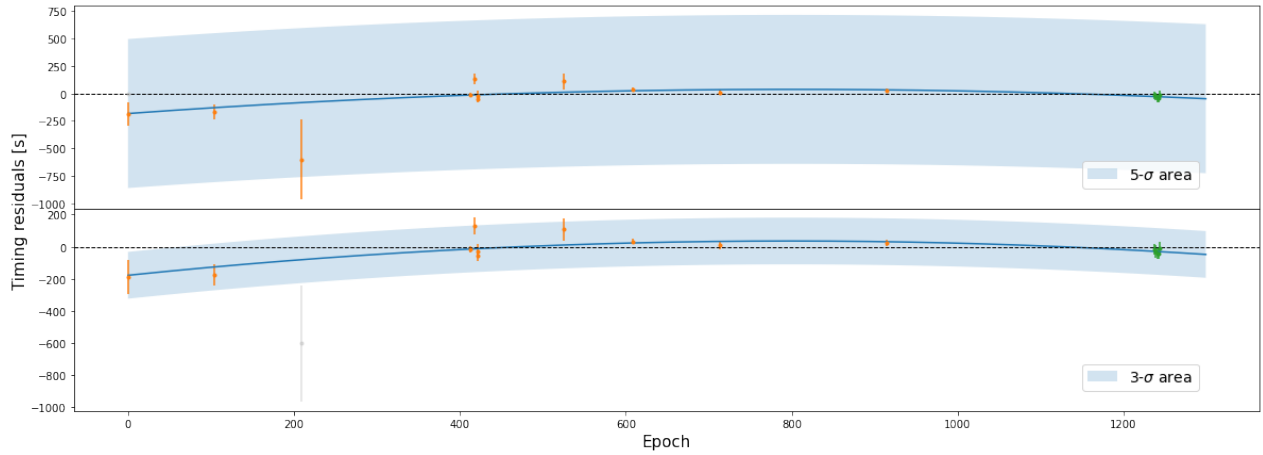

**Figure A8.** Timing residuals of WASP-22 b. The lines and symbols used are similar to Figure A1.

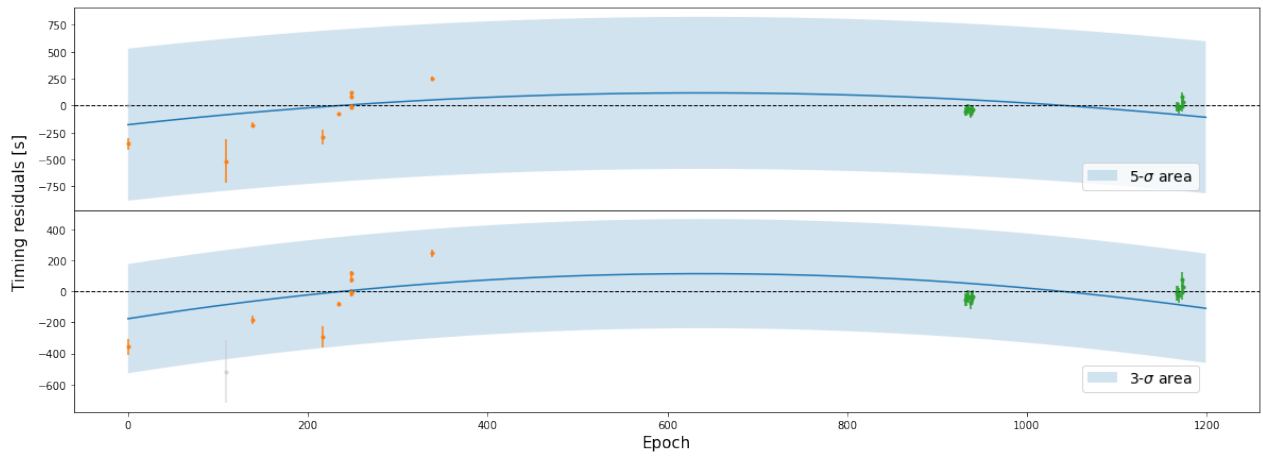

**Figure A9.** Timing residuals of WASP-45 b. The lines and symbols used are similar to Figure A1.

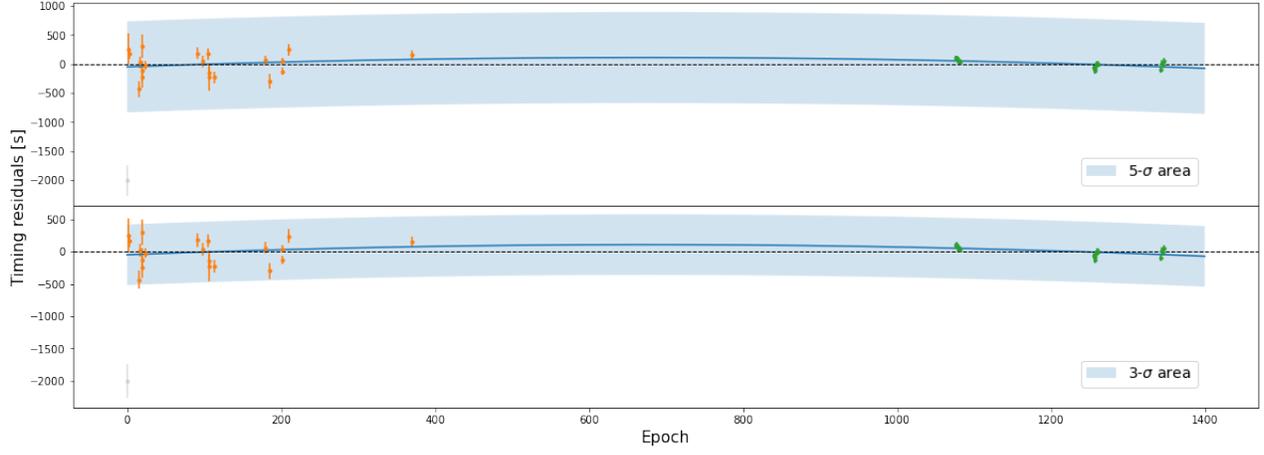

**Figure A10.** Timing residuals of XO-4 b. The lines and symbols used are similar to Figure A1.

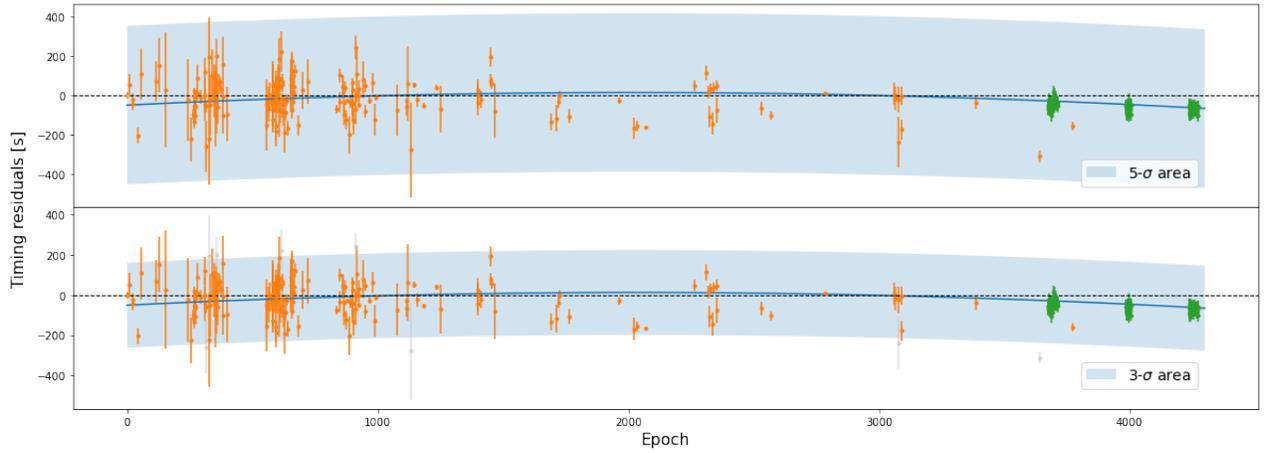

**Figure A11.** Timing residuals of TrES-3 b. The lines and symbols used are similar to Figure A1.

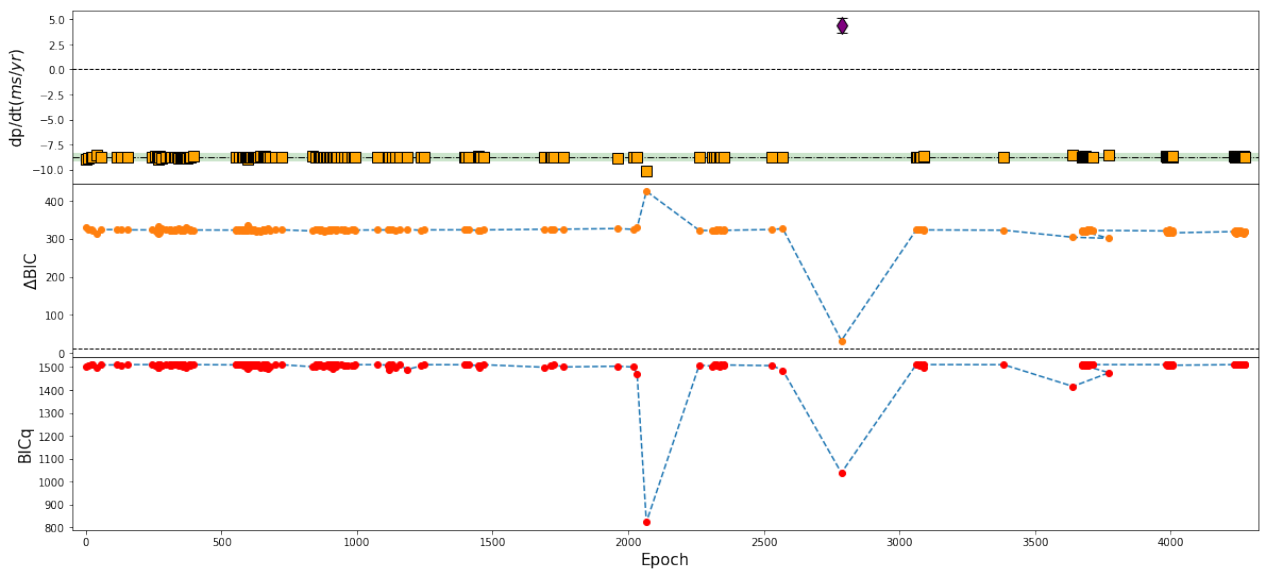

**Figure A12.** LOOCV analysis (top), corresponding ΔBIC (middle) and BICq (bottom) of TrES-3 b. The top panel displays the period change rate dp/dt obtained by fitting the quadratic model after the removal of each single transit timing data. The orange squares show the dp/dt values that satisfy the criterion of being $3\sigma$ away from zero, while the purple diamonds represent dp/dt values that fail to meet this criterion. The dash-dotted line marks the original best-fitting dp/dt value before the removal of any data, and the green shaded area marks the corresponding $1\sigma$ confidence region. The middle panel displays the corresponding ΔBIC. The bottom panel displays the BIC value from the quadratic ephemeris model fitting.

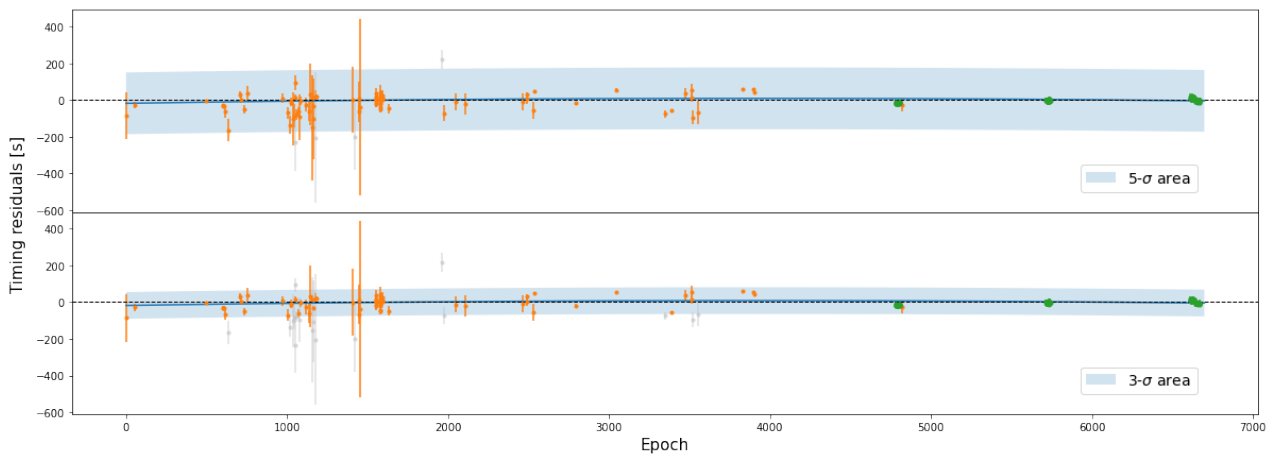

**Figure A13.** Timing residuals of WASP-19 b. The lines and symbols used are similar to Figure A1.

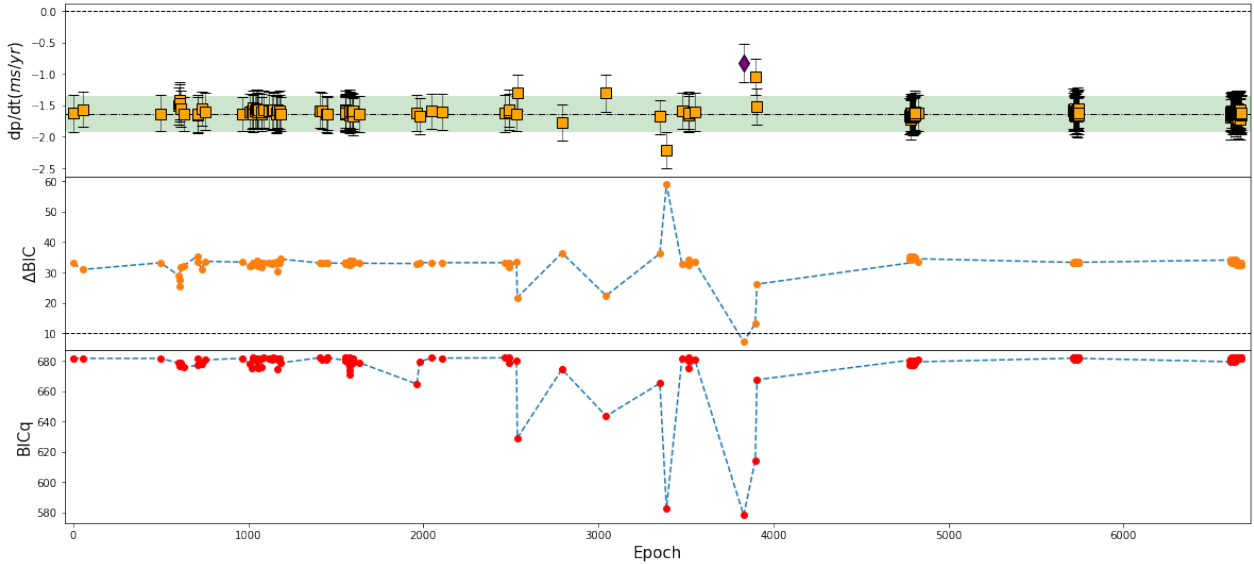

**Figure A14.** LOOCV analysis (top), corresponding ΔBIC (middle) and BICq (bottom) of WASP-19 b. The lines and symbols used are similar to Figure A12

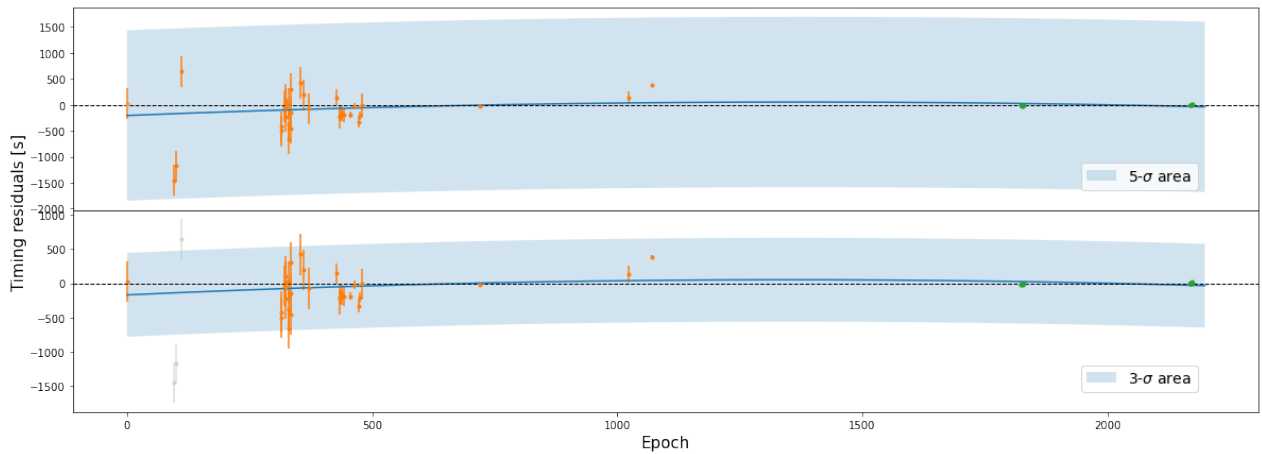

**Figure A15.** Timing residuals of XO-3 b. The lines and symbols used are similar to Figure A1.

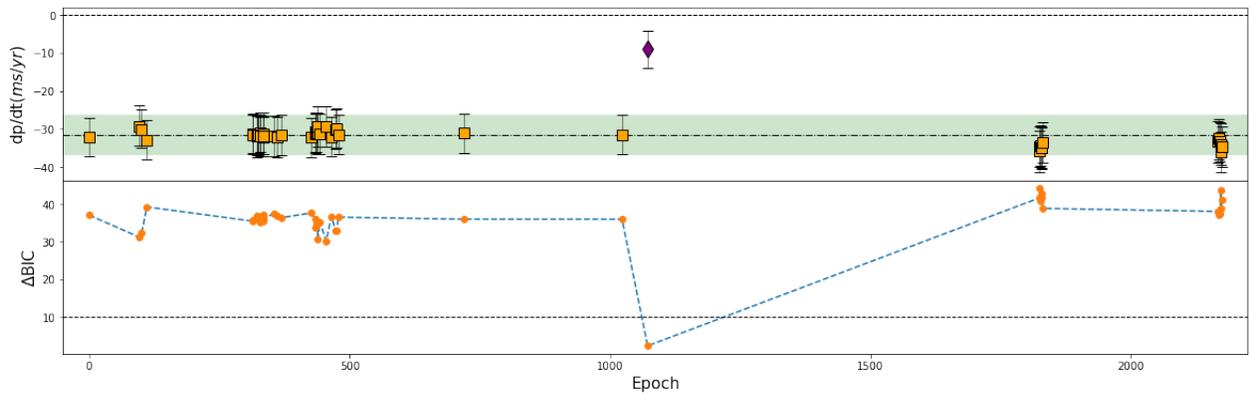

**Figure A16.** LOOCV analysis (top) and corresponding ΔBIC (bottom) for XO-3 b. The lines and symbols used are similar to Figure A12

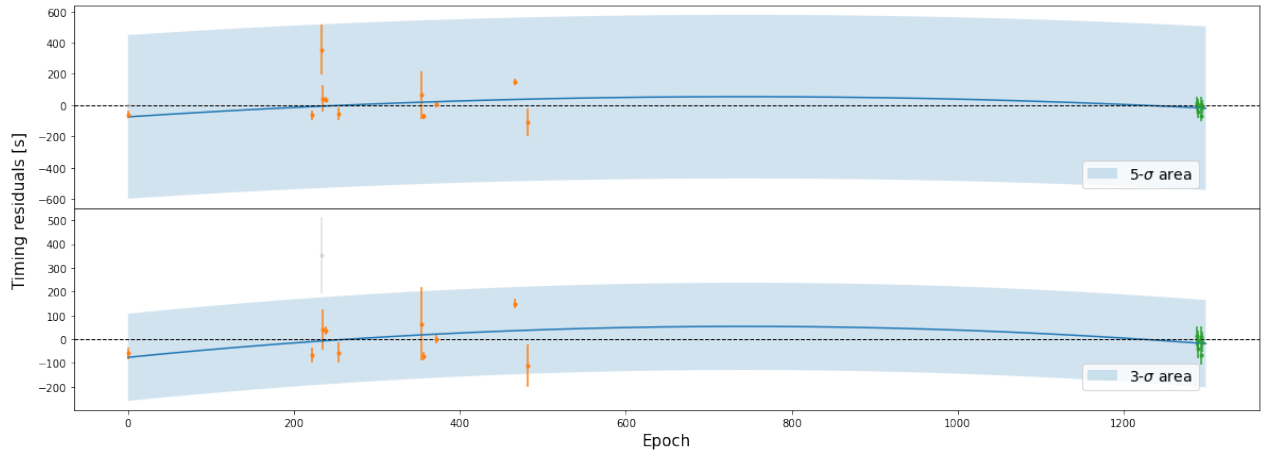

**Figure A17.** Timing residuals of WASP-16 b. The lines and symbols used are similar to Figure A1.

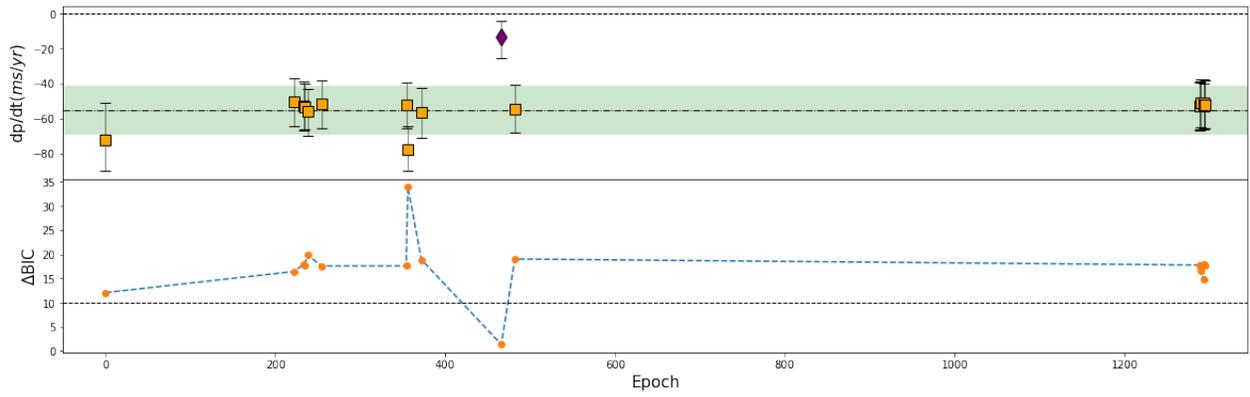

**Figure A18.** LOOCV analysis (top) and corresponding ΔBIC (bottom) of WASP-16 b. The lines and symbols used are similar to Figure A12

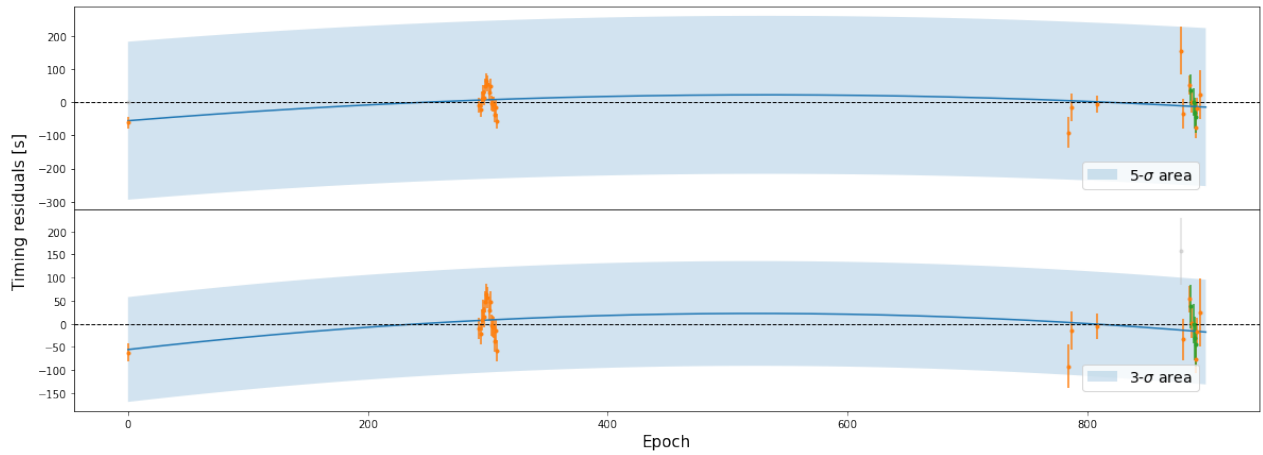

**Figure A19.** Timing residuals of WASP-47 b. The lines and symbols used are similar to Figure A1.

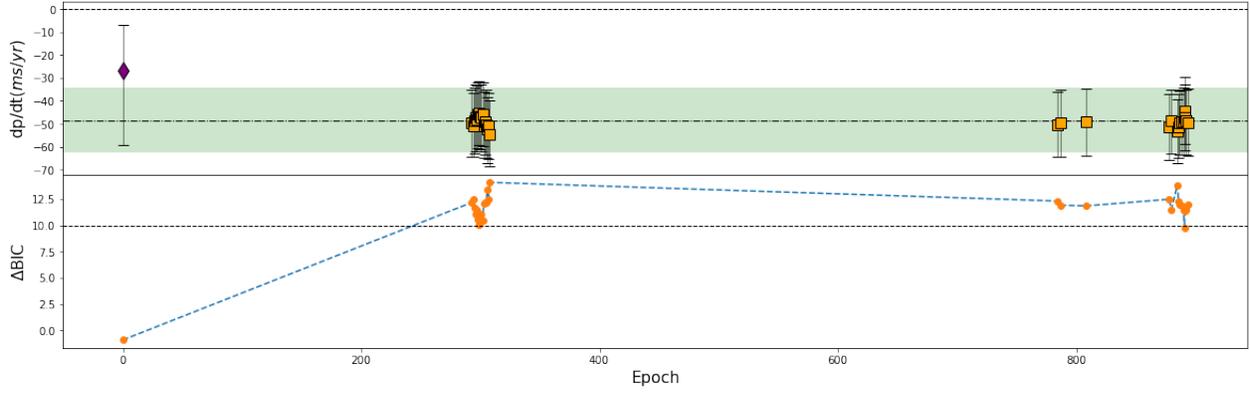

**Figure A20.** LOOCV analysis (top) and corresponding ΔBIC (bottom) of WASP-47 b. The lines and symbols used are similar to Figure A12

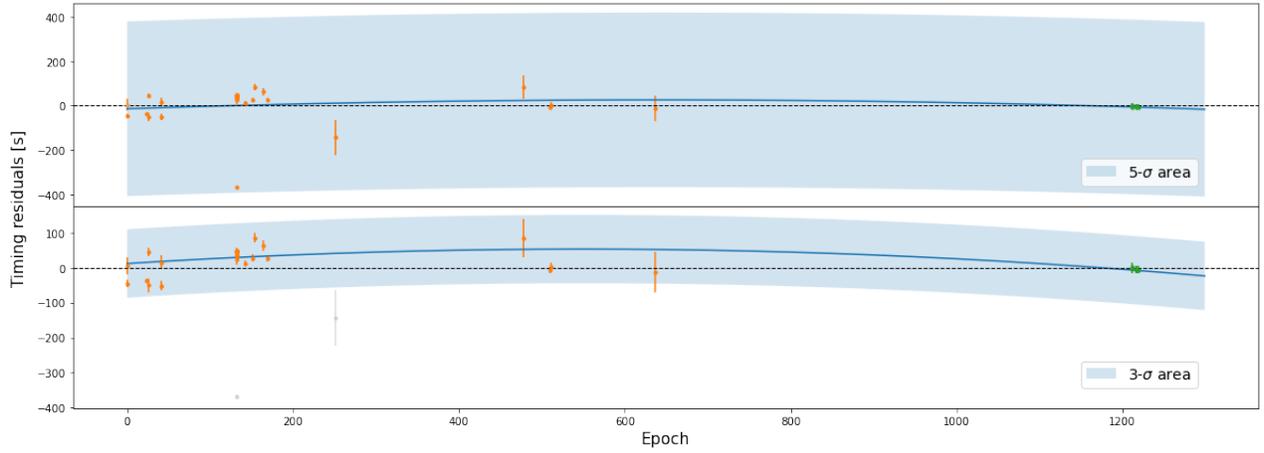

**Figure A21.** Timing residuals of WASP-80 b. The lines and symbols used are similar to Figure A1.

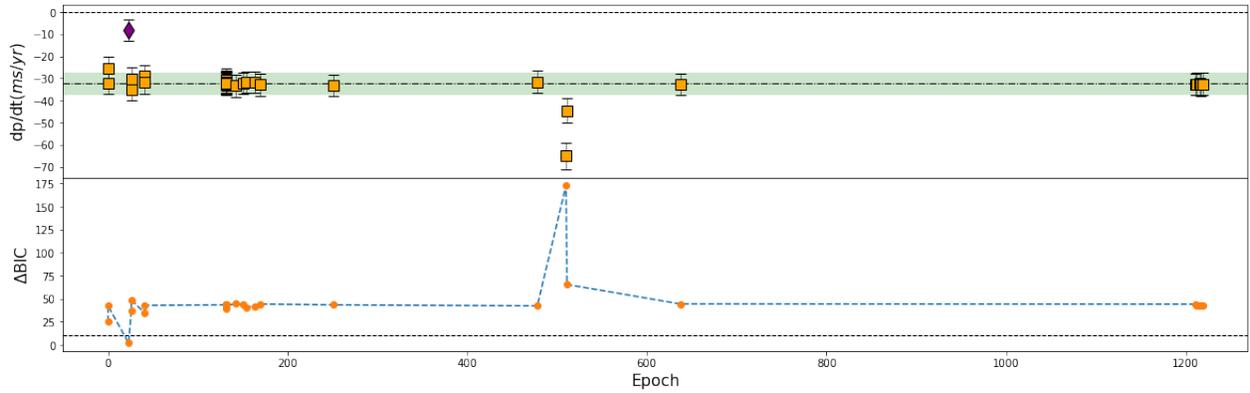

**Figure A22.** LOOCV analysis (top) and corresponding ΔBIC (bottom) of WASP-80 b. The lines and symbols used are similar to Figure A12

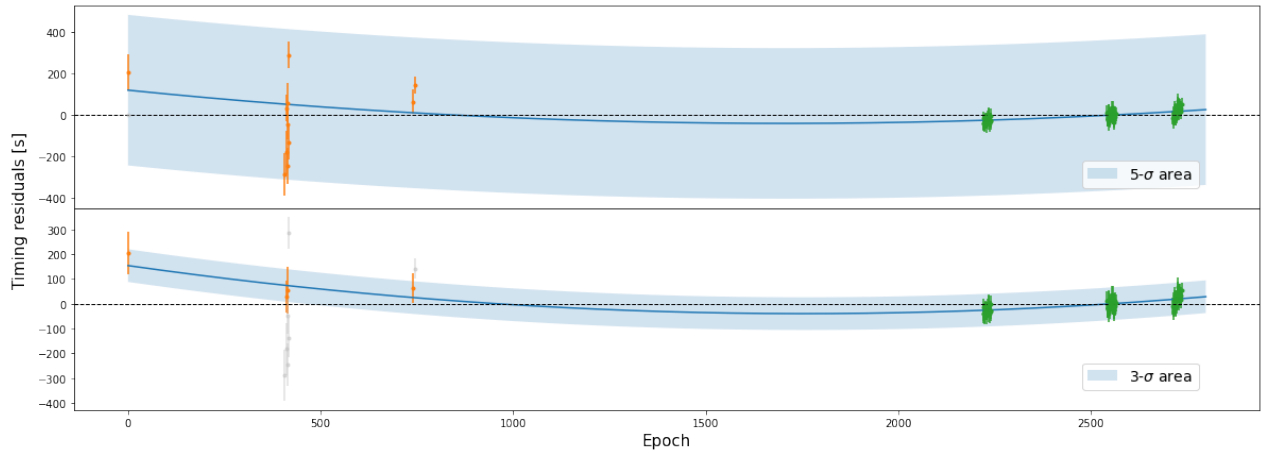

**Figure A23.** Timing residuals of HAT-P-7 b. The lines and symbols used are similar to Figure A1.

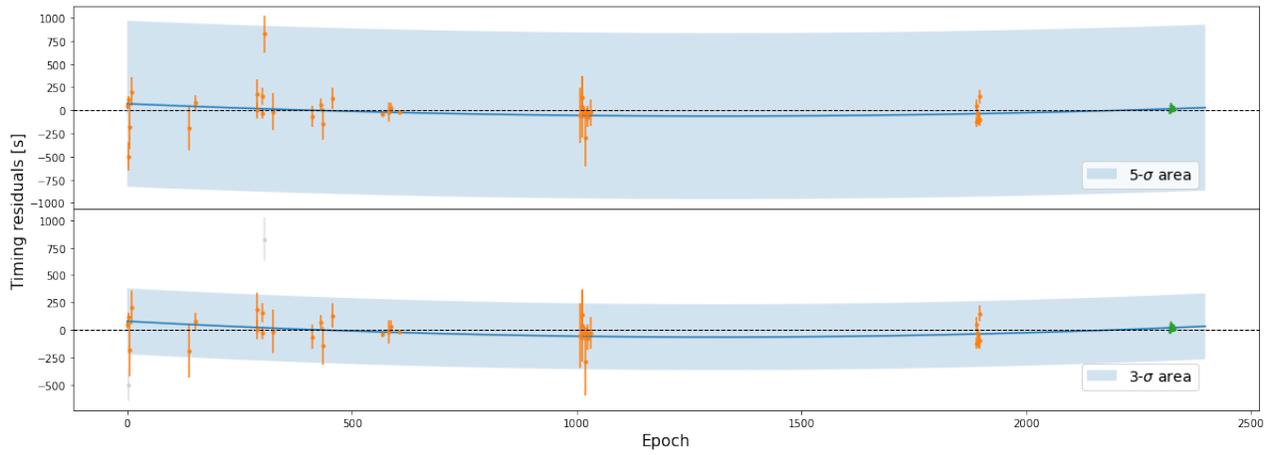

**Figure A24.** Timing residuals of WASP-1 b. The lines and symbols used are similar to Figure A1.

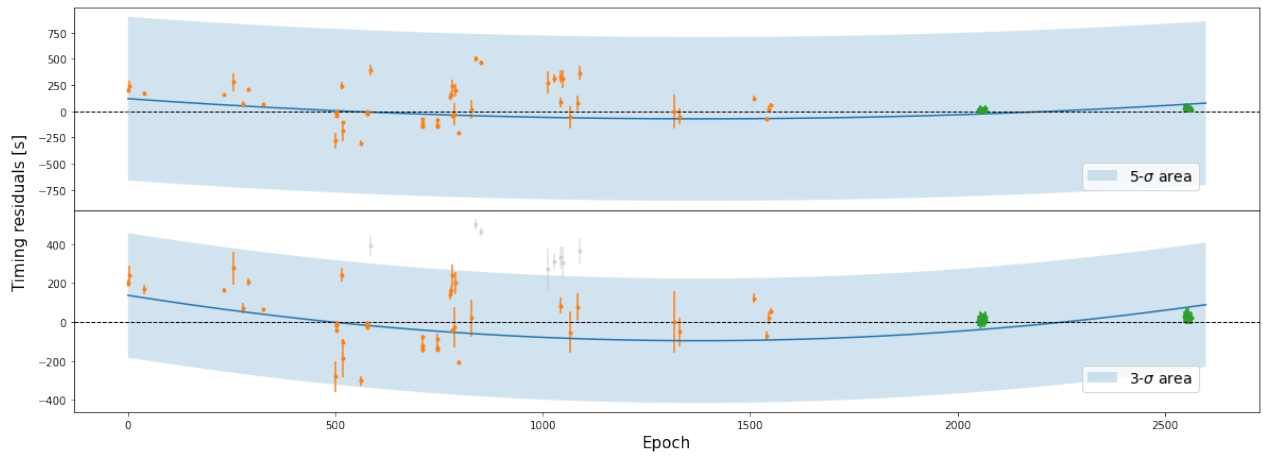

**Figure A25.** Timing residuals of WASP-46 b. The lines and symbols used are similar to Figure A1.

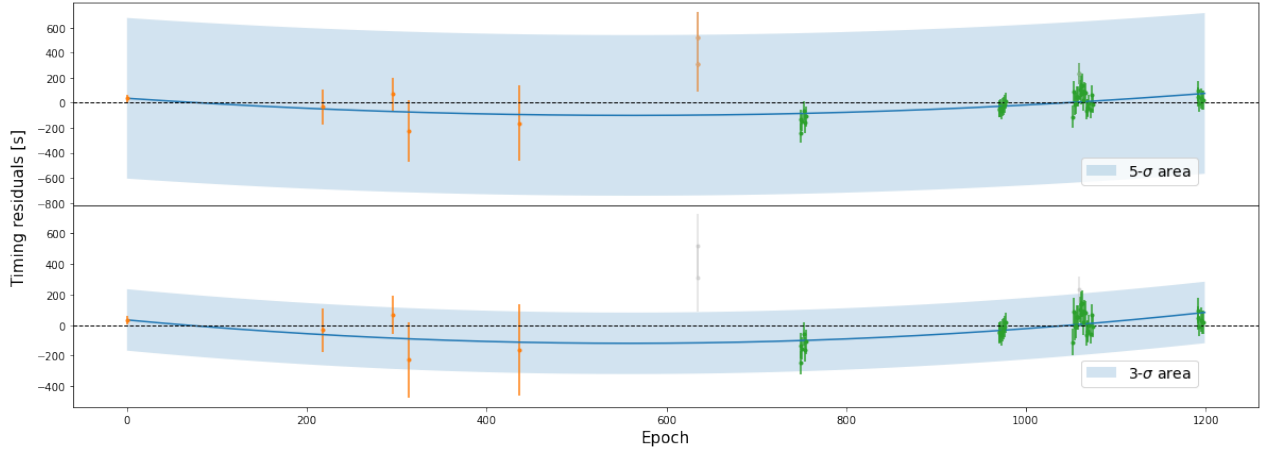

**Figure A26.** Timing residuals of HAT-P-43 b. The lines and symbols used are similar to Figure A1.

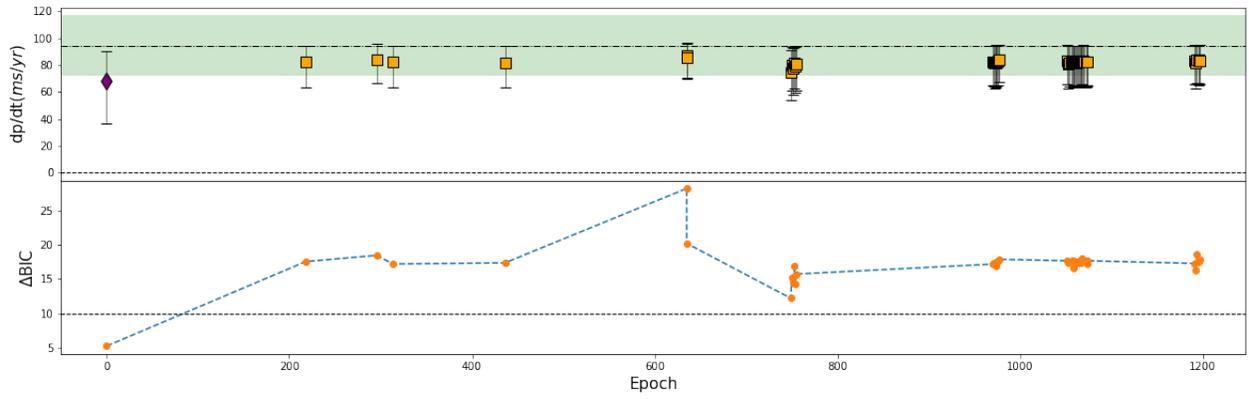

**Figure A27.** LOOCV analysis and corresponding ΔBIC of HAT-P-43 b. The lines and symbols used are similar to Figure A12.

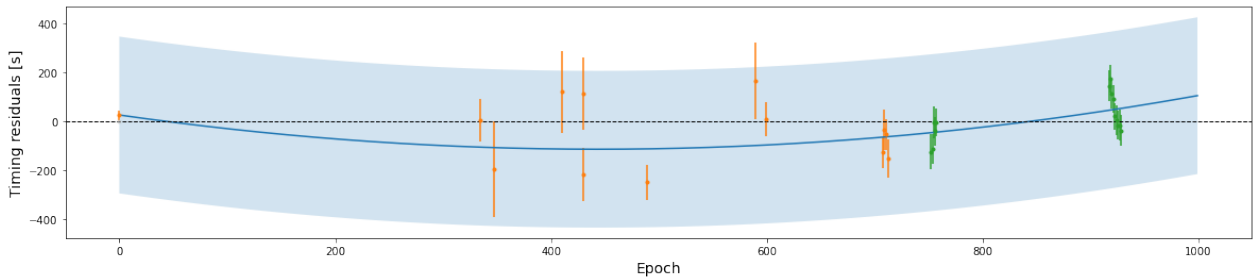

**Figure A28.** Timing residuals of HAT-P-44 b. The lines and symbols used are similar to Figure A1.

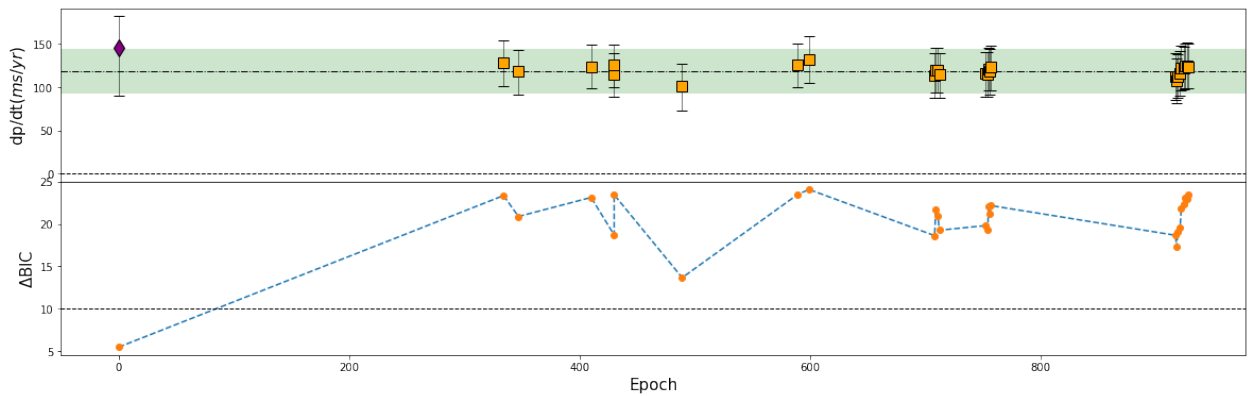

**Figure A29.** LOOCV analysis and corresponding ΔBIC of HAT-P-44 b. The lines and symbols used are similar to Figure A12.

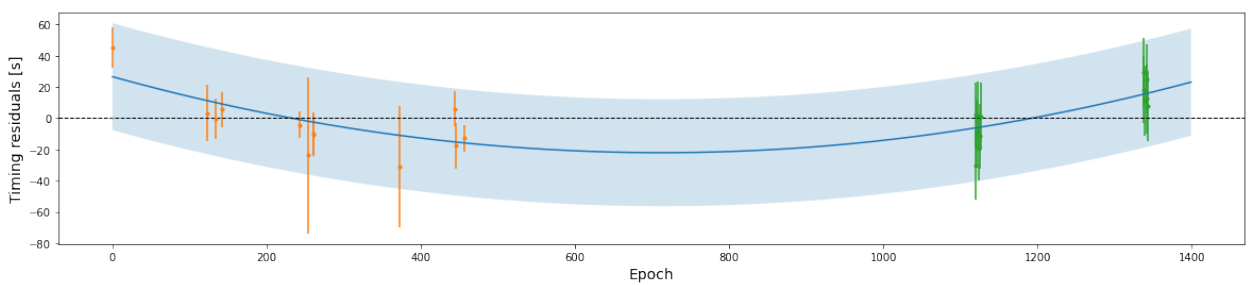

**Figure A30.** Timing residuals of WASP-6 b. The lines and symbols used are similar to Figure A1.

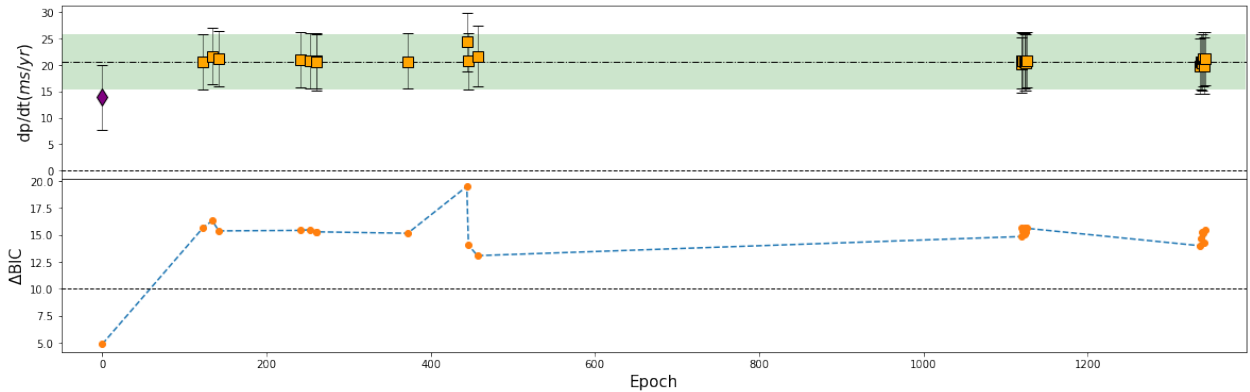

**Figure A31.** LOOCV analysis and corresponding ΔBIC of WASP-6 b. The lines and symbols used are similar to Figure A12.

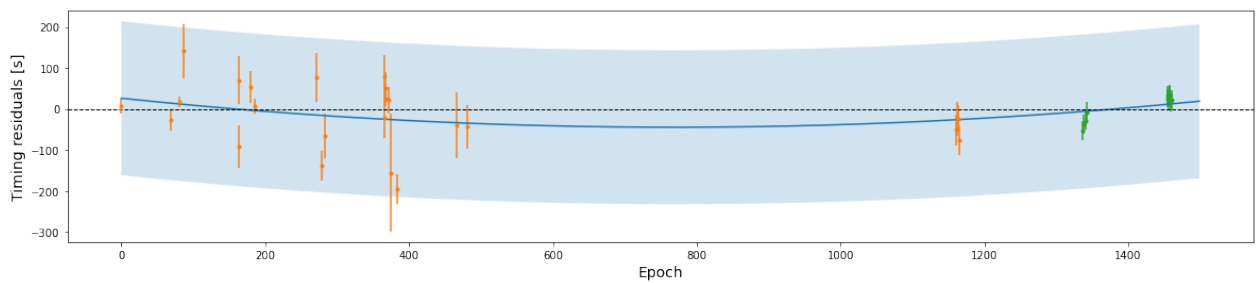

**Figure A32.** Timing residuals of WASP-11 b. The lines and symbols used are similar to Figure A1.

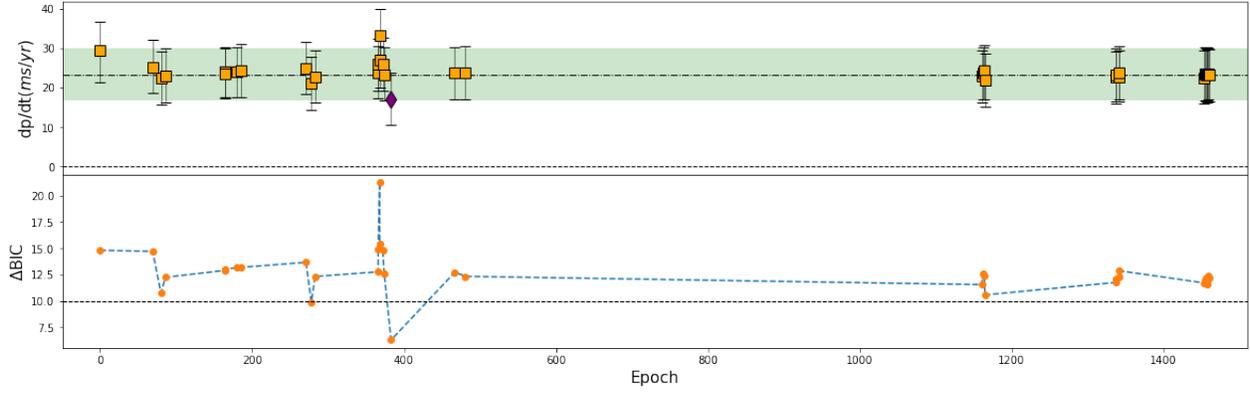

**Figure A33.** LOOCV analysis and corresponding $\Delta$BIC of WASP-11 b. The lines and symbols used are similar to Figure A12.

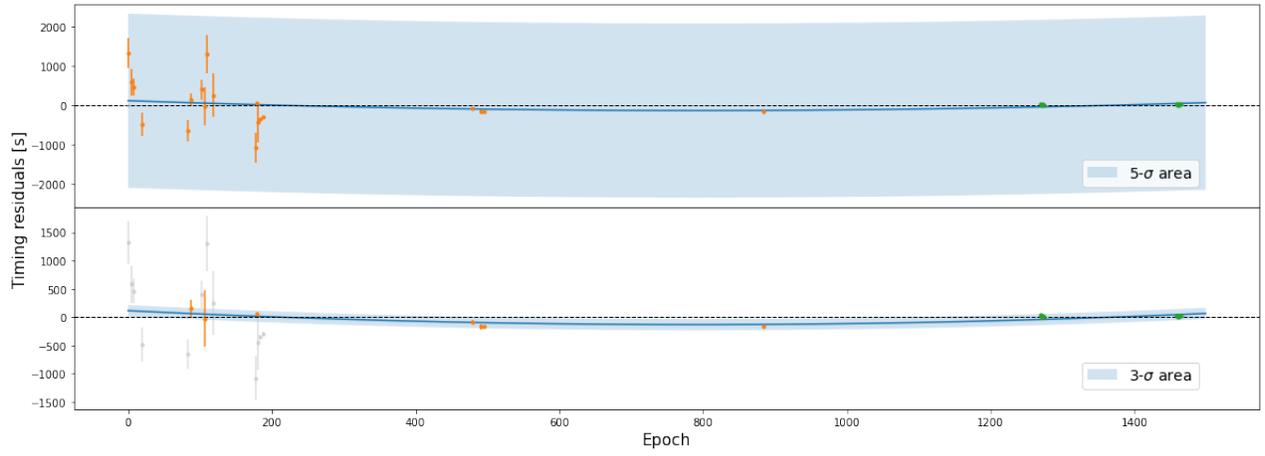

**Figure A34.** Timing residuals of WASP-17 b. The lines and symbols used are similar to Figure A1.

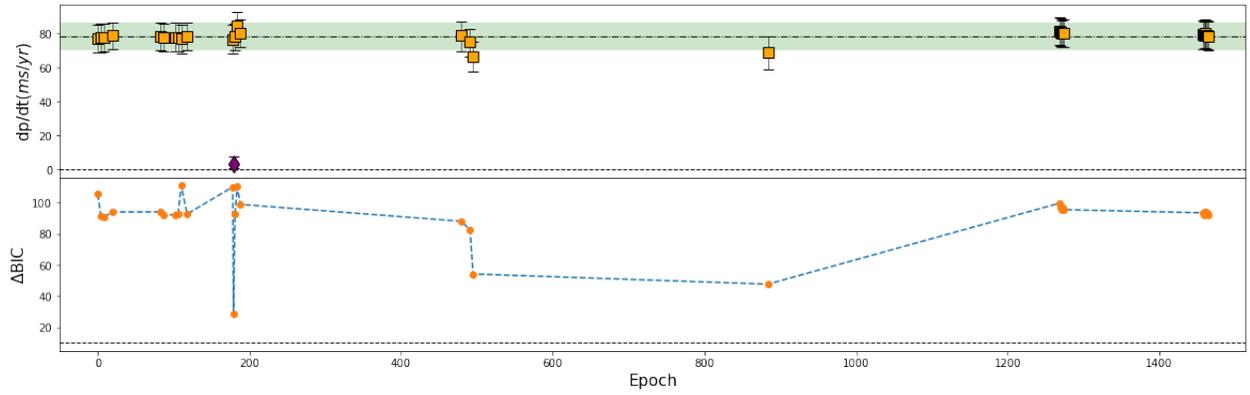

**Figure A35.** LOOCV analysis and corresponding $\Delta$BIC of WASP-17 b. The lines and symbols used are similar to Figure A12.